\documentclass[graybox]{svmult}
 \usepackage[utf8]{inputenc}

\usepackage{type1cm}        
\usepackage{makeidx}         
\usepackage{graphicx}        
\usepackage{multicol}        
\usepackage[bottom]{footmisc}

\usepackage{braket}
\usepackage{newtxtext}       %
\usepackage{newtxmath}       
\UseRawInputEncoding

\makeindex             

\title*{The Coming Decades of Quantum Simulation}
\begin{document}

\author{Joana Fraxanet, Tymoteusz Salamon and Maciej Lewenstein}
\institute{Joana Fraxanet \at ICFO - Institut de Ciencies Fotoniques, The Barcelona Institute of Science and Technology, 08860 Castelldefels (Barcelona), Spain. \email{joana.fraxanet@icfo.eu}
\and Tymoteusz Salamon \at ICFO - Institut de Ciencies Fotoniques, The Barcelona Institute of Science and Technology, 08860 Castelldefels (Barcelona), Spain. \email{tymoteusz.salamon@icfo.eu}
\and Maciej Lewenstein \at ICFO - Institut de Ciencies Fotoniques, The Barcelona Institute of Science and Technology, 08860 Castelldefels (Barcelona), Spain and ICREA, Pg. Llu\'{\i}s Companys 23, 08010 Barcelona, Spain. \email{maciej.lewenstein@icfo.eu}}
%

\maketitle
\tableofcontents

\newpage

\abstract{Contemporary quantum technologies face major difficulties in fault tolerant quantum computing with error correction, and focus instead on various shades of quantum simulation (Noisy Intermediate Scale Quantum, NISQ) devices, analogue and digital quantum simulators  and quantum annealers. There is a clear need and quest for such systems that, without necessarily simulating quantum dynamics of some physical systems, can generate massive, controllable, robust, entangled, and superposition states. This will, in particular, allow the control of decoherence, enabling the use of these states for quantum communications  (e.g. to achieve efficient transfer of information in a safer and quicker way), quantum metrology, sensing and diagnostics (e.g. to precisely measure phase shifts of light fields, or to diagnose quantum materials). In this Chapter we present a vision of the golden future of quantum simulators in the decades to come.}


\newpage

\section{Introduction and outline}
\label{sec:1}

This Chapter has a form of an essay on quantum simulators (QS) and represents personal opinions of the authors. It begins with Section \ref{sec:2}, which provides a brief review of the current status of quantum computing both in the scope of future fault tolerant quantum computers as well as Noisy Intermediate Scale Quantum (NISQ) devices, focusing on their general limitations. The content we present is based on the lectures and papers by many authors and which has been recently reviewed during ICFO Theory Lecture series on Quantum Computing by Alba Cervera-Lierta \cite{Alba,Alba1,Alba2}. 

Next, in Section \ref{sec:3}, we introduce the concept of quantum simulation, and discuss with some details and examples their: i) Ideology; ii) Platforms and architectures; and finally iii) New challenges for the coming decades. 

Sections \ref{sec:4} and \ref{sec:5} are devoted to a more detailed discussion of various achievements of quantum simulation. There, we discuss modern problems of physics that have been addressed in various systems. We divided them in two categories: i) Fundamental problems of physics and ii) Novel systems that exhibit novel physics. The former aims at utilizing quantum simulators as tools for better understanding of vital problems in condensed matter physics, high energy physics and quantum field theory. The latter addresses the progress of quantum simulation in the left-field and exotic areas such as ultra-fast processes, twisted multi-layer materials, strongly correlated phases of matter and Rydberg atoms. 

In Section \ref{sec:6}, we focus on novel methods in diagnostics and design of quantum many body systems, specific for QS and crucial for studying the phenomena mentioned above. In particular, we discuss: i) Detection with single site/single "particle" resolution; ii) Entanglement/topology characterization (with random unitaries); iii) Entanglement characterization with experiment-friendly approaches; iv) Topology characterization with experiment-friendly approaches; and last, but not least v) Synthetic dimensions. Finally, we list some of the new and old (but renewed) methods of theoretical physics that are being intensively developed in the age of QS. 
We conclude in Section \ref{sec:7}.
\section{Quantum Computing}
\label{sec:2}

\subsection{Classical computers and classical information processing} 

In classical computers, information processing is device independent -- the mechanical machines and supercomputers using electronics use the same principles; only the speed of calculation and memories are different. The basic information processing unit in classical information theory \cite{Cover,McKay} is a BIT, which can take two values:
             $$|0\rangle\;  {\rm or}\; |1\rangle.$$

Classical computers have limitations: there exist difficult, if not impossible, tasks for them. Computer scientists classify problems according to complexity classes \cite{wiki-cc}. The simplest ones are $L$ and $NL$, defined as the class of problems solvable with logarithmic (time or space) resources on a deterministic (non-deterministic) Turing machine, respectively. Similarly $P$ and $NP$ require polynomial time resources on deterministic (non-deterministic) Turing machines. On the other extreme, there are problems that are proven to require exponential resources in time and/or space, $EXPTIME$ and $EXPSPACE$. 
While it is known that
   $$L \subseteq NL\subseteq P \subseteq NP \subseteq PSPACE \subseteq EXPTIME \subseteq EXPSPACE,$$
it is not known if the inclusions are strict: one of the most challenging problems of computer science is whether $P=NP$. 

Another thing worth mentioning is that "device independence" of classical computation is also an illusion. As first discussed by Rolf Landauer, computation is a physical process that costs energy, work, and heat. In fact, a lot of these resources: contemporary supercomputers pay astronomic costs for cooling the installations. Landauer's ideas are summarized in his famous Landauer's Principle, according to which erasing a bit in a system of temperature $T$ costs $k_BT/2$, where $k_B$ is the Boltzmann constant \cite{Landauer}. 

\begin{figure}[t]
\centering
\includegraphics[width=5cm]{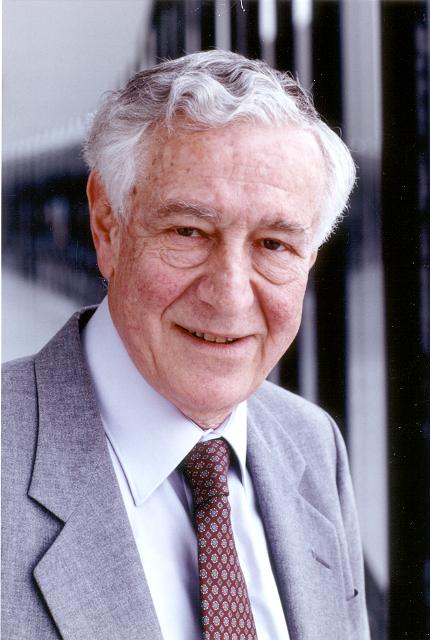}
\caption{Rolf Landauer (1927-1999). This photo is taken from https://ethw.org/File:landauer.jpg, reprinted with permissions of IEEE History Center.}
\label{fig:1}       
\end{figure}

\subsection{Quantum information processing} 

According to the father of Quantum Information (QI) science, Charles H. Bennett, information processing in QI is apparently strongly device dependent (quantum mechanics decides about the laws of information processing). 

\begin{figure}[t]
\centering
\includegraphics[width=7cm]{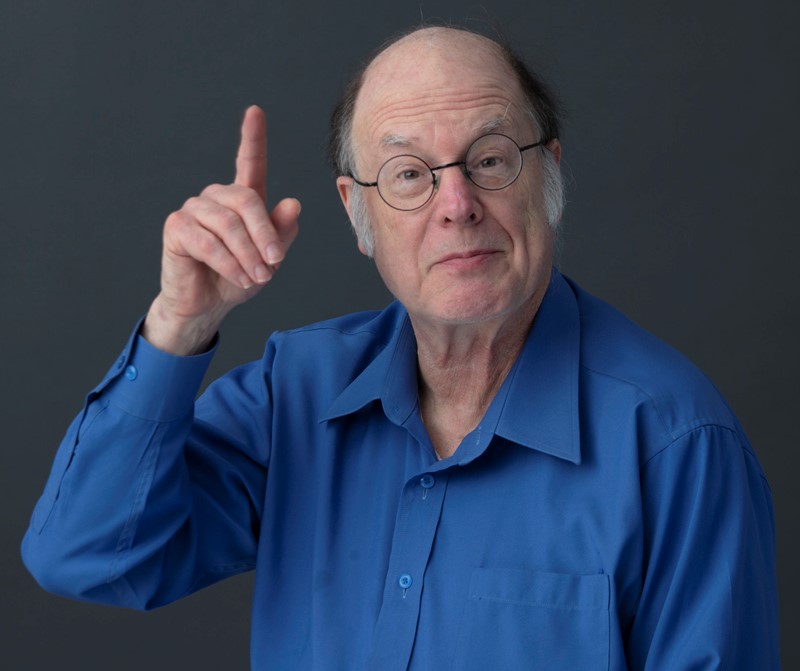}
\caption{
Charles H. Bennet (born 1943). This photo was taken on February 28, 2018. $\copyright$ SmugMug+Flickr/IBM Research/51002548905 under CC BY-SA 2.0.}
\label{fig:2}       
\end{figure}

The basic information processing unit in QI is a QUBIT, which may
attain superposition states
    $$ \alpha|0\rangle  +  \beta|1\rangle.$$
                                 
Quantum computers are intrinsically parallel and may be much faster than classical ones; but they also have have limitations: quantum super-positions live short, due to interactions with the external world.

\subsection{Universal quantum computers} 

A universal quantum computer realizes arbitrary unitary operators $U$  ("isometry") acting on the vector space of $N$ qubits (qutrits, qudits, ...). The device acts in a space which is exponentially large: the so called Hilbert space has (for qubits) dimension $2^N$. A quantum computer can be regarded as a quantum circuit. According to Wikipedia: "In quantum information theory, a quantum circuit is a model for quantum computation, similar to classical circuits, in which a computation is a sequence of quantum gates, measurements, initializations of qubits to known values, and possibly other actions. The minimum set of actions that a circuit needs to be able to perform on the qubits to enable quantum computation is known as DiVincenzo's criteria \cite{DiVincenzo}." 
   
A universal quantum computer requires a universal set of 
quantum gates. A possible universal set contains all single qubit ones and the so called CNOT gate, which entangles two qubits. 
Calculations correspond to applying quantum gates -- single qubit gates or two qubit gates -- but also to applying quantum measurements. 

Unfortunately, calculations in quantum computers are very fragile and require fault tolerant error correction. This is similar to classical machines, but it is much more difficult to realize with quantum devices. For instance, to have one logical qubit we need an overhead of $10^4$ additional qubits to ensure error correction (\@$10^3$ errors for the typical, so called, surface codes). Thus, a quantum computer with $100$ logical qubits requires $10^6$ qubits to function well.

Moreover, error scaling should be such that the error per gate is independent of the size of the device and of the number of times the gates are used, but this is not the case in the presently available devices \cite{Alba}. Obviously, these requirements  are much beyond current technological feasibility, and constitute a true scientific and technological challenge.

\begin{figure}[t]
\includegraphics[scale=.375]{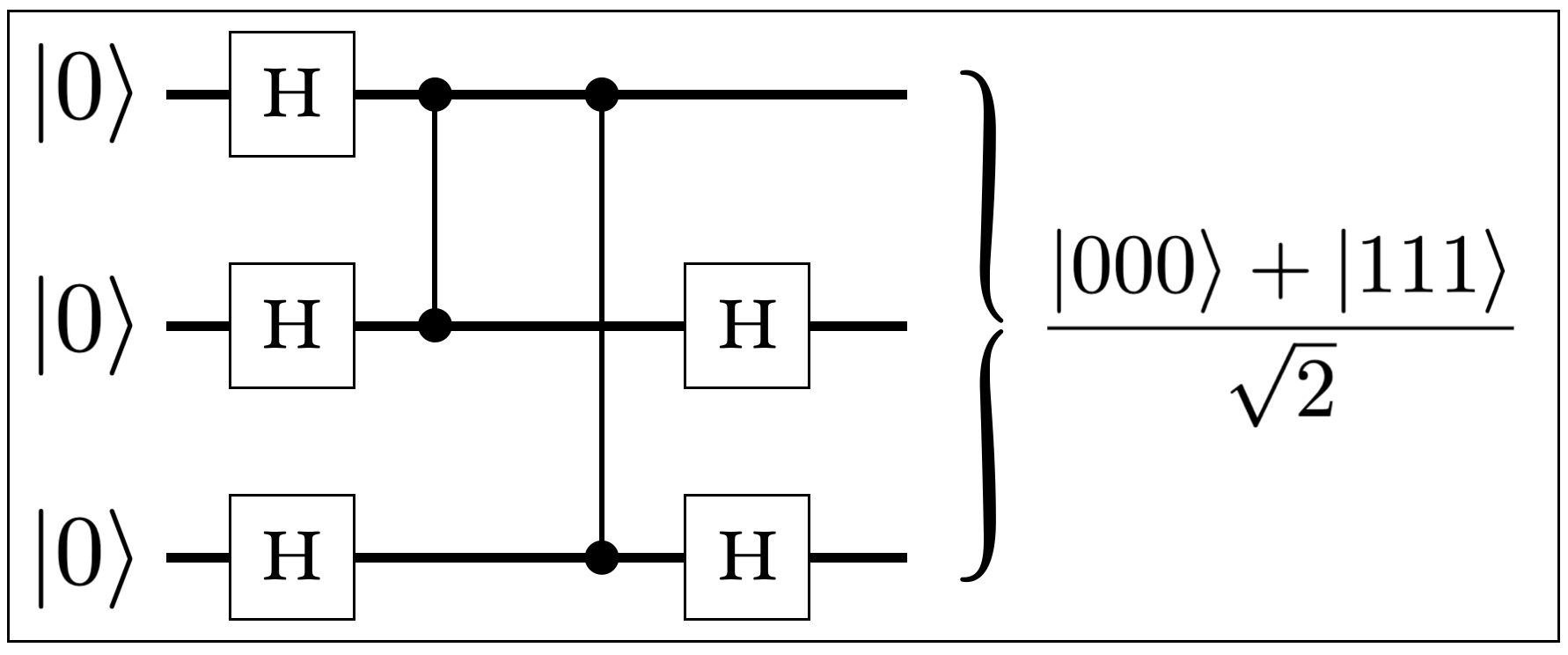}
\caption{Example of a quantum circuit}
\label{fig:3}       
\end{figure}

\subsection{Noisy Intermediate Scale Quantum devices}

All of this stimulated John Preskill \cite{Preskill18} to propose seeking for quantum advantage with Noisy Intermediate Scale Quantum (NISQ) devices, which: i) operate with 50-100 "dirty" qubits without fault tolerant error correction; ii) still, are capable of show some advantage over classical computers. So far, NISQ devices have been used to: 

\begin{itemize}                                                                                                
\item Demonstrate {\bf Quantum Supremacy} in systems that have been rigorously proven to potentially provide advantage by quantum computer scientists (cf. \cite{Aaronson}). So far, two experiments claim such advantage: an experiment by Google with superconducting qubits \cite{Google} and an experiment by the group of Jianwei Pan with photons \cite{JWPan}. Both essentially perform sampling (the outcome measurements) from random unitary circuits. In the case of Josephson junction superconducting qubits, the random unitary circuits are nonlinear, while in the case of boson sampling in a photonic circuit they are linear. Three things should be stated clearly: i) the experiments are beautiful and mark amazing progress in experimental capabilities; ii) the demonstrated quantum advantage concerns purely academic problems without having technological applications so far; iii) it is not yet clear whether the advantage is true, i.e., we do not know how far we can optimize classical algorithms that describe sampling problems, although we know that they are $NP$-hard.

\item Realize alternative forms of quantum computing, such as {\bf Adiabatic Quantum Algorithms}
\cite{Farhi} and/or {\bf Quantum Annealing} \cite{D-Wave}. In both cases: i) the NISQ devices are used as special purpose quantum computers, {\it ergo} quantum simulators; ii) so far, the quantum advantage in the rigorous sense is, delicately  speaking, disputable (cf. \cite{Biswas,KatzgraberTroyer}). 

\item Realize {\bf Quantum Variational Eigensolvers}.
Quoting the recent review \cite{Tilly}: "The variational quantum eigensolver (or VQE) uses the variational principle to compute the ground state energy of a Hamiltonian, a problem that is central to quantum chemistry and condensed matter physics. Conventional computing methods are constrained in their accuracy due to the computational limits. The VQE may be used to model complex wavefunctions in polynomial time, making it one of the most promising near-term applications for quantum computing. Finding a path to navigate the relevant literature has rapidly become an overwhelming task, with many methods promising to improve different parts of the algorithm. Despite strong theoretical underpinnings suggesting excellent scaling of individual VQE components, studies have pointed out that their various pre-factors could be too large to reach a quantum computing advantage over conventional methods." 

\item Realize {\bf Quantum Approximate Optimization Algorithms (QAOA)}, which, according to Wikipedia, are quantum algorithms that are used to solve optimization problems. "Mathematical optimization deals with finding the best solution to a problem (according to some criteria) from a set of possible solutions. Mostly, the optimization problem is formulated as a minimization problem, where one tries to minimize an error which depends on the solution: the optimal solution has the minimal error. The power of quantum computing may allow problems which are not practically feasible on classical computers to be solved, or suggest a considerable speed up with respect to the best known classical algorithm." 
Typically, QAOA are formulated as hybrid algorithms, where the quantum processor calculates the cost functions (cf. energy, energy gradients, etc.) which depend on variational parameters, and the classical computer performs a classical optimization of these parameters. Again, according to the recent Nature Physics \cite{Patron}: "The bounds we obtain indicates that substantial quantum advantages are unlikely for classical optimization unless noise rates are decreased by orders of magnitude or the topology of the problem matches that of the device. This is the case even if the number of available qubits increases substantially." 

\item Realizing {\bf Quantum Machine Learning}. Also in this application the prospects and even reason for searching quantum advantage is questionable, as discussed in the recent paper of Maria Schuld and Nathan Killoran \cite{Schuld}: "In this perspective we explain why it is so difficult to say something about the practical power of quantum computers for machine learning with the tools we are currently using. We argue that these challenges call for a critical debate on whether quantum advantage and the narrative of "beating" classical machine learning should continue to dominate the literature the way it does, and provide a few examples for alternative research questions."

\end{itemize}

As Sankar Das Sarma writes in \cite{sankar}: "Quantum computing has a hype problem. Is quantum computing thus just a hype? No, it is a fair and serious effort, but realistically practical quantum supremacy of importance for science and technology remains in the domain of special purpose quantum computers, i.e. {\bf quantum simulators}". In particular, there already exists IBM quantum computing infrastructure that people can access and run their quantum algorithms there.  In fact, there have been quite a few of applications in chemistry and materials run in that platform with interesting results, not yet at the point of overruling what can be done with classical computers. It is worth noticing that recently many of the cases of the claimed quantum supremacy \cite{Google} have been beaten by contemporary classical simulations using tensor networks \cite{Pan1,Pan2}. Also, the prospects for exponential  quantum supremacy in quantum chemistry are questionable \cite{Garnet}. The very near future will bring surely answers to some these questions.

\section{Quantum Simulators}
\label{sec:3}

\subsection{Idea and main concepts} 

A simulator is a device that imitates desired properties of a given system such as an airplane, train or a physical system of quantum particles. Airplane simulator cannot fly, nor take passengers on board but provides realistic reproduction of all navigational, mechanical and electronic apparatus necessary for the efficient and safe training of the crew. Similarly, simulators of the quantum systems do not reproduce all properties of the original, simulated systems, but imitate only their specific features that are required to understand the phenomena of interest occurring in the original system. The general idea and the concept of such quantum simulators (QSs) can be shortly sketched as follows: 
\begin{itemize}
    \item There exist many interesting {\bf quantum phenomena} with highly important applications(such as, for instance, superconductivity).
   \item These phenomena are often complex to describe and understand with the help of standard or even super computers.
   \item Maybe we can design another, simpler and more controllable quantum system to simulate, understand and control these phenomena, as proposed originally by Yu. I.  Manin and R. P. Feynman \cite{Manin,Feynman}. Such a system would thus work as a quantum computer of one specific purpose, i.e. a {\bf quantum simulator}. 
\end{itemize}
Designing such a simulator is however not an easy task at all. The research line on QS goes back to the beginning of this century and there are numerous valuable reviews covering the various platforms and types of QS's (cf. \cite{Nature}). In fact, the beginning of the practical concept to QS goes back to the proposal for simulating strongly correlated systems in optical lattices \cite{Bose-Hubbard} and the first experiments \cite{BlochHansch}. Nowadays, QS are commonly used for the following tasks and goals: 
\begin{itemize}
     \item {\bf Fundamental problems of physics.} This is the most developed application, in which many achieved results are believed to reach quantum advantage; this is particularly true for the studies of quantum dynamics, or quantum disordered systems, such as the ones that exhibit many body localization (MBL). 
     \item {\bf Quantum chemistry.} Applications of quantum NISQ devices and QS to quantum chemistry has only started \cite{Babbush,npj,Alba,Guzik} and, although promising, it is still far from achieving the precision and accuracy of contemporary theoretical quantum chemistry.
     There is a growing evidence that the expect of exponential quantum advantage in quantum simulations for quantum chemistry is not such (see recent works by G.K. Chan \cite{Chan}, and his talks at the APS meeting in 2022). 
     
     \item {\bf Classical/quantum optimization problems for technology.} Applications of quantum NISQ devices and QS to optimization problems are also in an initial phase \cite{QAOA} and cannot yet compete with the classical supercomputer methods (cf. \cite{Patron}).
\end{itemize}

\subsection{Platforms and architectures}  
Now is the time to mention how the quantum simulators can be actually built. There are several very well developed platforms where QSs are being realized and "practical" quantum advantage has been achieved. We should stress that QSs can be analog or digital.

In the latter case, any platform that offers tools for universal quantum computing can also be used for quantum simulation. Here comes a definitely incomplete list of already developed QS architectures: 

\begin{itemize}
    \item {\bf Superconducting qubits}: These are the same  systems as the ones used by Google \cite{Google} or D-Wave \cite{D-Wave}. Even though in principle they allow for "noisy" but universal quantum computing, they can be and very often are used as digital QSs (cf. \cite{Parao}). They can be coupled to microwave cavities, resulting into circuit QED systems \cite{Wallraff}.
	\item {\bf Ultracold atoms}: They mostly offer the possibility for analog quantum simulation. This can be realized in the continuum or in optical lattices \cite{LSA2017}. They are very flexible and they allow to simulate complex Hubbard models, as well as spin systems. 
	\item {\bf Trapped ions}: Similarly to superconducting qubits, trapped ions allow for universal quantum computing, but they caalso be used as perfect analog or digital QSs \cite{Monroe53,MonroeRMP}. Similarly to superconducting qubits, they can be used to simulate spin $1/2$ systems, rather than Hubbard models. Very recently a qudit  (or spin $1$ and $2$) quantum computer/simulator was realized with ions \cite{Ringbauer}.
	\item  {\bf Rydberg atoms}: These are atoms where the electron has been excited to a high principal quantum number, and which are trapped in optical tweezers. They mimic spin systems with long range interactions \cite{Lukin51,LukinScars,Browaeys}.
	\item {\bf Photonic systems}: These are typically linear optics systems which, combined with photon counting, may mimic a universal quantum computer, according to the famous paper from Knill, Laflamme, and Milburn \cite{KLM}. Achieving strong non-linearity with photons is very challenging, but there are ongoing attempts and proposals \cite{Shahaf}.
	\item {\bf Light and Cavity materials}: Quantum Simulators based on Cavity Quantum Electrodynamics take advantage of the coupling between quantum system and the coherent light field of the cavity, in which such system has been placed. This branch of quantum simulation is commonly named as "cavity quantum materials", since one in principle could place a many-body quantum system into a cavity and control its properties via light-matter interactions. Currently, however, the experimental studies are mainly conducted in the scope of Jaynes – Cummings and Dicke models \cite{Schlawin_2022}. Another research path was taken by engineering materials entirely from light with resulting photon-photon interactions\cite{Clark_2020,carusotto_2020,Ma_2019,Schine_2016}. Such systems, however, require a  mediator (for example Rydberg-dressed atoms)  facilitating the light-light interactions.  
	\item {\bf Twistronics systems}: Twistronics deals with twisted bilayer graphene or other two-dimensional materials \cite{Pablo,Dima}. For small "magic" angle, such systems lead to periodic moir\'e patterns at a length scale much larger than the typical scale of condensed matter systems: in this sense, they can themselves be considered as condensed matter quantum simulators of condensed matter \cite{Kennes21}. Such approach has been explained in detail in Chapter 1 of this book. Twisted bilayer materials can, however, also be mimicked by ultracold atoms in a two-dimensional lattice with synthetic dimensions \cite{tymek}.
	\item {\bf Polaritons}: Especially useful for non-equilibrium systems and quantum hydrodynamics simulation, as well as relativistic effects thanks to dual (half light half particle) nature of the polaritonic quasi-particles \cite{Basov2021, hubener2021,JBloch}.
\end{itemize}

 \subsection{The coming decades of quantum simulation} 
 
 In the coming decades, as clearly reflected in the current and future quantum programs such as National Quantum Initiative Act, Quantum Flagship, MIC 2025 and many others, the platforms that we use are not expected to change, but the challenges and focus are going to be very different. QSs of the future will have to be devices that are: 
 \begin{itemize}
 {\bf    \item Robust
      \item Scalable
      \item Programmable
      \item Externally accessible
      \item Standardized
      \item Verifiable and certifiable}
 \end{itemize}
 
Moreover, the priorities concerning the main goals will be revised, increasing the effort in optimization problems with application to technology, which we predict to be the main axis of interest. Quantum chemistry will most likely remain on the second place of the podium. Fundamental problems of physics will not loose importance, however they will most probably no longer be the unique problems where quantum advantage can be achieved. Below we highlight the main (in our opinion) efforts in each of above mentioned areas that will be undertaken in the upcoming decades.

\begin{itemize}
      \item {\bf Classical/quantum optimization problems for technology.} Quantum simulation is already starting to address practically relevant and computationally hard problems, such as the Maximum Independent Set problem, which was very recently addressed with Rydberg atom arrays \cite{Lukin2022}. We expect this type of applications to very strongly developed in the coming years.  
      \item {\bf Quantum chemistry.} This will for sure include novel methods of simulating quantum chemistry going beyond NISQ devices to analog simulators (cf. \cite{Tudela}).
      \item {\bf Fundamental problems of physics.} Besides the simulation of Hamiltonian systems for condensed matter and high-energy physics, the generation, manipulation and applications of massively entangled states, which useful for quantum communication, quantum metrology, sensing and detection, will gain momentum. In the coming years, we also expect QSs to focus on particularly complex and classically hard to simulate problems, including the Fermi-Hubbard model and the puzzle of high-temperature superconductivity, the simulation of lattice gauge theories and related problems in dynamical lattices (for reviews see \cite{Bañuls,Royal}) or frustrated and disordered systems, among others.
\end{itemize}

One should stress that currently, there is a huge effort led I. Bloch in Munich to make this a user service infrastructure in the short term (“sort of providing quantum computing access using atoms in optical lattices as quantum platform). In an even more general sense, the Quantum Flagship project PASQuanS - Programmable Atomic Large-Scale Quantum Simulation combines all the efforts for atomic systems: Rydberg atoms, trapped ions, ultracold atoms on optical lattices and traps, and more (https://pasquans.eu/).

\section{Quantum simulation of fundamental problems of physics}
\label{sec:4}

In this Section, we present a selected list of some of the great achievements of QS over the last few years. Typically, we quote and describe some experimental papers from the leading groups over the years, and also present also some examples from our own groups. This Section is, by no means, a review. It is instead based on subjective choices and selections (for a recent review see \cite{RecentUSA}).

\subsection{Relevant paradigmatic systems}

\paragraph{\textit{Fermi-Hubbard model}} 

A "paradigmatic" and notoriously difficult to classically simulate systems is, of course, the Fermi-Hubbard model, believed by many to lie at heart of high-temperature superconductivity \cite{PALee}. In recent years, major achievements in the quantum simulation of the Fermi-Hubbard model have been reported by the groups of Markus Greiner at Harvard and Immanuel Bloch in Munich, which are using quantum gas microscopes to explore antiferromagnetically ordered systems and the physics that emerges when perturbing them through doping. A good example of what QS can do in this context is the paper from the Harvard group \cite{Strings}, where the authors write: "Understanding strongly correlated quantum many-body states is one of the most difficult challenges in modern physics. For example, there remain fundamental open questions on the phase diagram of the Hubbard model, which describes strongly correlated electrons in solids. In this work we realize the Hubbard Hamiltonian and search for specific patterns within the individual images of many realizations of strongly correlated ultracold fermions in an optical lattice. Upon doping a cold-atom antiferromagnet we find consistency with geometric strings, entities that may explain the relationship between hole motion and spin order, in both pattern-based and conventional observables. Our results demonstrate the potential for pattern recognition to provide key insights into cold-atom quantum many-body systems." Interestingly, machine learning approaches can be exploited to analyze some of these experiments, cf. \cite{MLGreiner}.

\paragraph{\textit{Systems with fine-tuned interactions: supersolids and quantum liquid droplets}} 

\begin{figure}[t]
\centering
\includegraphics[width=5cm]{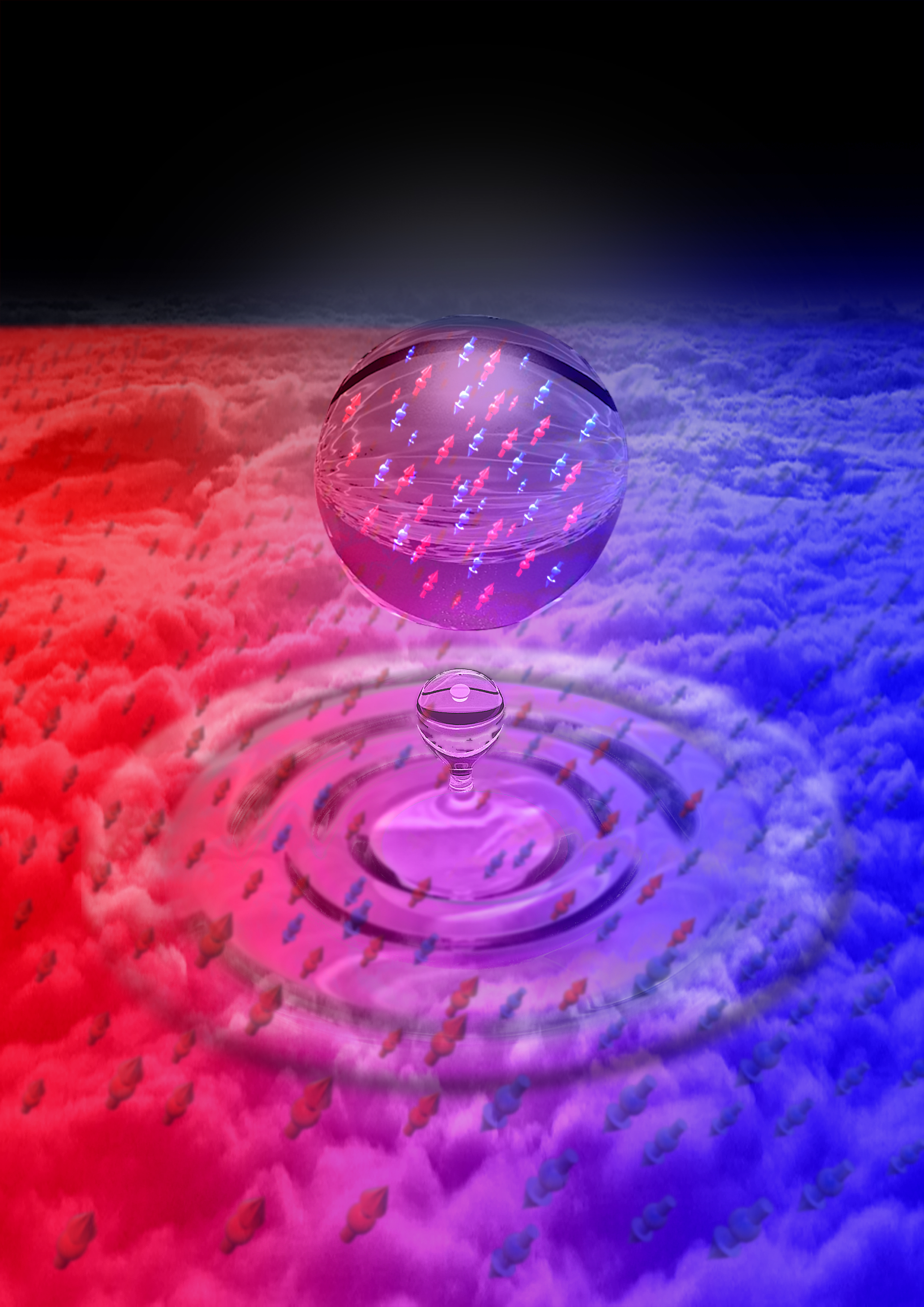}
\caption{Artistic view of a quantum liquid droplet formed by mixing two Bose-Einstein condensates with competing repulsive intrastate and attractive interstate interactions \cite{CabreraScience2018}. Reprinted with permission of ICFO/PovarchikStudiosBarcelona.} 
\label{fig:leticia}       
\end{figure}

Another recent research direction in quantum simulation is the study of systems where interactions of different origins compete. When they have opposite signs, such interactions can be fine-tuned to a situation where they practically compensate each other. The resulting systems can then become completely dominated by quantum fluctuations and correlations despite remaining weakly interacting, leading to the emergence of novel phases of matter such as supersolids or quantum liquid droplets \cite{BoettcherRoPP2021}.

Supersolids are systems that combine the frictionless flow of a superfluid and the crystalline structure of a solid. Despite being predicted in the '50s, they remained elusive until quantum gases made their QS possible, using either atoms dressed with light or dipolar quantum gases with competing interactions. Quantum liquid droplets, on the other hand, are ensembles of quantum particles that are hold together by attractive interparticle interactions, but stabilized against collapse by the repulsive effect of quantum fluctuations, and which have been observed in dipolar Bose gases and in mixtures of Bose-Einstein condensates (see Fig. \ref{fig:leticia}).

\subsection{Fundamental systems of condensed matter, high energy physics and quantum field theory}

In the recent years, there has been a strong focus on QS of high energy physics (HEP) models, including lattice gauge theories (LGTs) and related models \cite{Bañuls,Royal}. The challenge here is to control many body interactions: "magnetic" interactions on the plaquettes of the lattice, and "electric" interactions, that are linked to matter through the local conservation laws of the gauge theory --  the Gauss law -- which ensures its local gauge invariance. 

\paragraph{\textit{Schwinger model}}

Due their versatility, trapped ions are particularly suitable to simulate LGTs in a digital way. These systems may serve as universal quantum computers and realize, in principle, any few body interactions and constraints. Here we quote two achievements of Rainer Blatt's group, where one-dimensional LGTs are implemented. In Ref. \cite{Blatt1}, they "report the first experimental demonstration of a digital quantum simulation of a lattice gauge theory, by realising $1+1$-dimensional quantum electrodynamics (Schwinger model) on a few-qubit trapped-ion quantum computer.  They are interested in the real-time evolution of the Schwinger mechanism, describing the instability of the bare vacuum due to quantum fluctuations, which manifests itself in the spontaneous creation of electron-positron pairs. To make efficient use of the quantum resources, they map the original problem to a spin model by eliminating the gauge fields in favour of exotic long-range interactions, which have a direct and efficient implementation on an ion trap architecture. They explore the Schwinger mechanism of particle-antiparticle generation by monitoring the mass production and the vacuum persistence amplitude. Moreover, they track the real-time evolution of entanglement in the system, which illustrates how particle creation and entanglement generation are directly related. Their work represents the first step towards quantum simulating high-energy theories with atomic physics experiments, the long-term vision being the extension to real-time quantum simulations of non-Abelian lattice gauge theories."

In the following paper \cite{Blatt2}, the authors consider hybrid classical-quantum algorithms that aim at solving optimisation problems variationally, using a feedback loop between a classical computer and a quantum co-processor, while benefiting from quantum resources. They "present experiments demonstrating self-verifying, hybrid, variational quantum simulation of lattice models in condensed matter and high-energy physics. Contrary to analog quantum simulation, this approach forgoes the requirement of realising the targeted Hamiltonian directly in the laboratory, thus allowing the study of a wide variety of previously intractable target models. Here, they focus on the lattice Schwinger model, a gauge theory of 1D quantum electrodynamics. Their quantum co-processor is a programmable, trapped-ion analog quantum simulator with up to 20 qubits, capable of generating families of entangled trial states respecting symmetries of the target Hamiltonian. They determine ground states, energy gaps and, by measuring variances of the Schwinger Hamiltonian, they provide algorithmic error bars for energies, thus addressing the long-standing challenge of verifying quantum simulation."

\paragraph{\textit{Bosonic Schwinger model}}

\begin{figure}[t]
\centering
\includegraphics[width=\columnwidth]{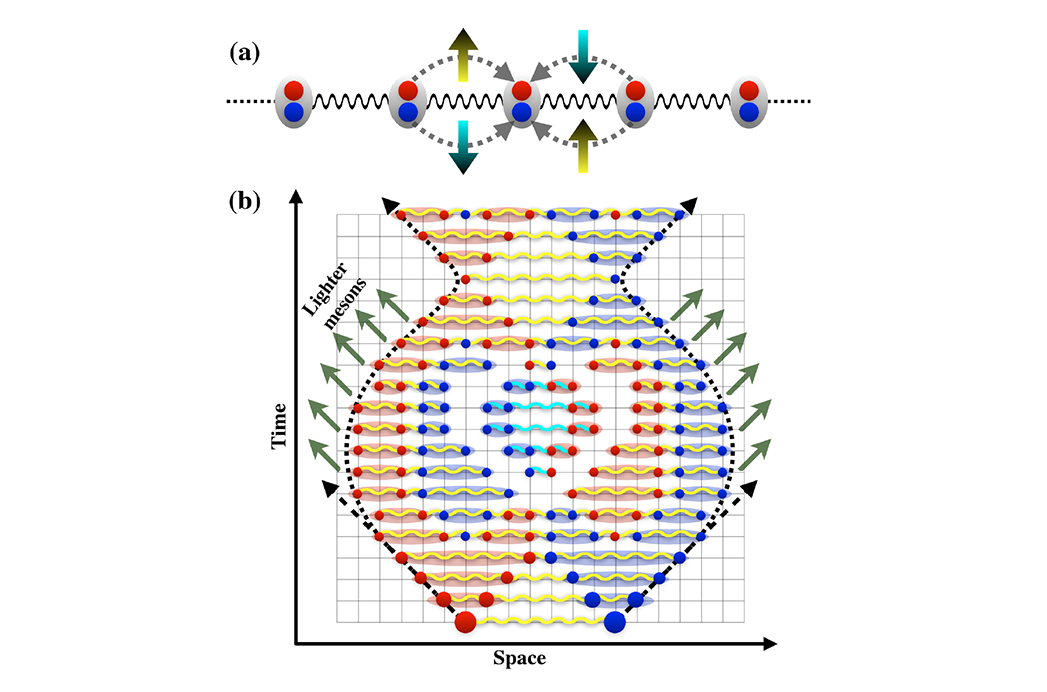}
\caption{(a) In the lattice version of bosonic Schwinger Model (BSM), sites are occupied by two kinds of bosons, corresponding to particles (red dots) and antiparticles (blue dots), which are coupled to U(1) gauge fields residing on the bonds of the lattice. Tunneling of bosons across a given bond change the internal state of the corresponding gauge field as dictated by the arrows. (b) The time-dependent dynamics of the BSM is greatly affected by strong confinement, where the light-cone of the particles bends inwards, producing exotic asymptotic states made of a central core of strongly correlated bosons and external regions populated by free mesons. Figure's author: Titas Chanda.} 
\label{fig:titas1}       
\end{figure}

Relativistic quantum gauge theories are fundamental theories of matter describing nature. Paradigmatic examples are quantum electrodynamics (describing electromagnetic interactions of charged particles and photons), chromodynamics (describing strong interactions of quarks and gluons), and the Standard Model, unifying the latter two with the weak interactions.

Despite enormous progress in our understanding of quantum gauge theories, the questions concerning the behaviors of systems described by such theories in the presence of strong correlations remain widely open: from the very nature of the quark confinement to the behavior of quark-gluon-plasma at high densities and temperatures. Moreover, quantum out-of-equilibrium dynamics of quantum gauge theories is out of reach of the present classical computers. For these reasons, there is a lot of effort to design and investigate quantum simulators of such systems. The paradigmatic model of quantum gauge theories in one spatial dimension and time is the Schwinger model, in which "charged" electrons (fermions) interact with photons (bosons) in one dimension. Since quantum simulations with fermions are notoriously difficult, simulating a bosonic version of the Schwinger model is an interesting and more accessible goal \cite{Titas1}.

Using state-of-the-art methods of theoretical physics, Titas Chanda and co-workers investigated in their work how the bosonic matter behaves when it is driven out-of-equilibrium by creating a pair of particle and antiparticle on top of the vacuum of the system. They obtain three important results for the understanding of quantum gauge theories in general: i) the bosons undergo an evolution which is dominated by strong confinement of charges, responsible for only a partial screening of electric field, even in the massless limit; ii) the extended "meson" formed by the two charges and the electric-flux tube connecting them is very robust, leaving a strong footprint in the entanglement of the system; iii) the system fails to thermalize and generates exotic states at long times, characterized by two distinct space-time regions -- one external region made by thermal mesons, and a central region between the two initial charges, where quantum correlations obey the, so called, area-law of entanglement (see Fig. \ref{fig:titas1}). This work opens a path towards quantum simulations of quantum gauge theories in novel, unexplored regimes.

\paragraph{\textit{Abelian-Higgs mechanism}}

\begin{figure}[t]
\centering
\includegraphics[width=9cm]{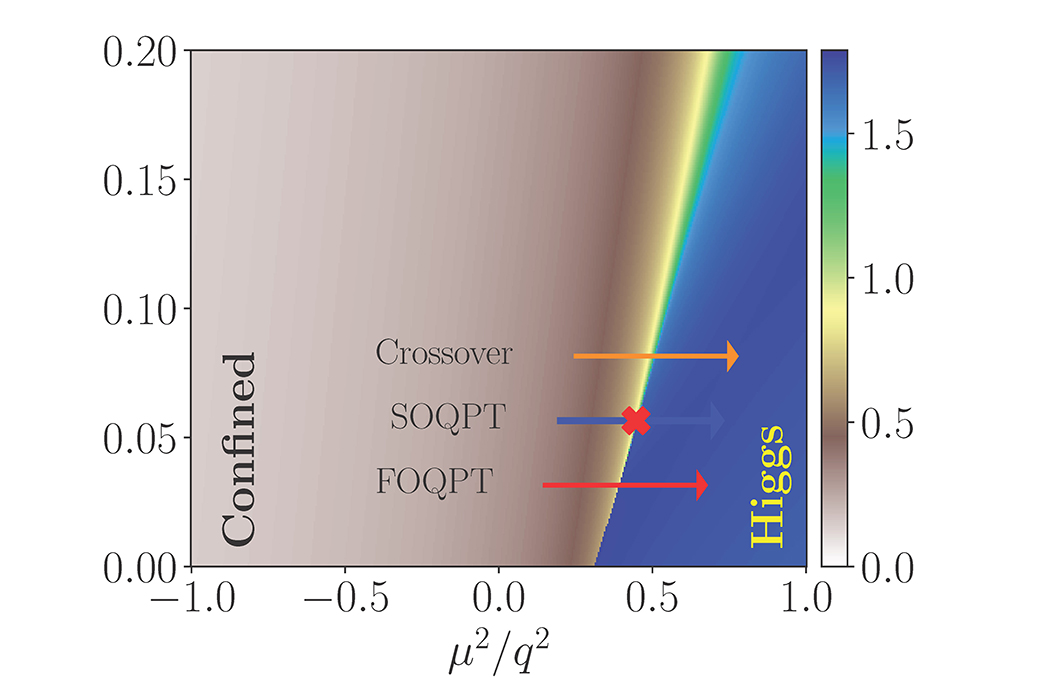}
\caption{Phase diagram of the lattice Abelian-Higgs model as seen through the lens of entanglement entropy. At small couplings, the system occupies two qualitatively different regions, a confined and a Higgs region, separated by a line of first order quantum phase transitions. Figure's author: Titas Chanda.} 
\label{fig:titas2}       
\end{figure}

The current focus of quantum simulation of fundamental models of high energy physics is lattice gauge theories and quantum field theories. The Higgs mechanism is an essential ingredient of the Standard Model of particle physics that explains the `mass generation' of gauge bosons. Its seemingly simple one-dimensional lattice version may serve as an interesting novel quantum simulator.

In a recent study \cite{Titas2}, a continuation of \cite{Titas1},  we have taken up the challenge to fill this gap. Unlike the system in the continuum, two distinct regions in the lattice version are identified, namely the confined and Higgs regions. These two regions are separated by a line of first order phase transitions that ends in a second order critical point. Above this critical point the regions are smoothly connected by a crossover. The presence of a second order critical point allows one to construct an unorthodox continuum limit of the theory that is described by a conformal field theory (CFT) (see Fig. \ref{fig:titas2}).
This work is strongly motivated by the current prospects of quantum simulation of quantum gauge theories, and it opens a path towards observing the Higgs mechanism in experiments with ultracold atomic setups.

\section{Novel quantum simulators for novel physics (NOQIA)}
\label{sec:5}

In this Section, we mention several novel directions/platforms for QS: i) QS for strong laser-matter interacting ii) twistronics, which can be regarded as condensed matter QS of condensed matter, and photonic moir\'e `patterns; iii) QS for strongly correlated phases of matter, which include indirect excitons and interaction induced topological order and iv) Rydberg atoms, which can simulate very interesting spin models with long-range interactions.

\subsection{Quantum simulation based on atto-science and ultra-fast processes}

Recently, we have proposed a completely new platform for QS based on atto-science and ultra-fast processes. Over the past four decades, astounding advances have been made in the field of laser technologies and the understanding of light-matter interactions in the non-linear regime. Thanks to this, scientists have been able to carry out extremely complex experiments related to, for example, ultra-fast light-pulses in the visible and infrared range, and accomplish crucial milestones such as using a molecule's own electrons to image its structure, to see how it rearranges and vibrates or breaks apart during a chemical reaction.

\paragraph{\textit{Optical Schr\"odinger cat states}}

\begin{figure}[t]
\centering
\includegraphics[width=7cm]{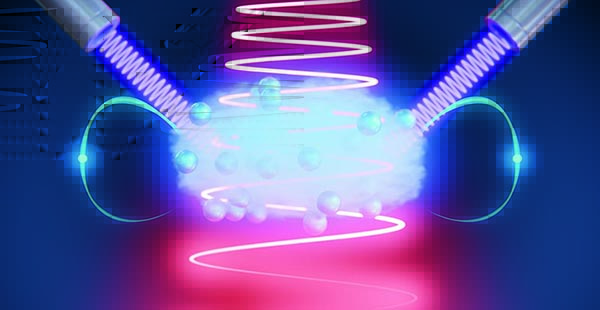}
\caption{Optical cat states in intense laser-matter interaction. Reprinted with permission of ICFO / Scixel - E. Sahagun.} 
\label{fig:paris}       
\end{figure}

The development of high-power lasers allowed scientists to study the physics of ultra-intense laser-matter interactions which, in its standard version, treats ultra-strong ultra-short driving laser pulses only from a classical point of view. The famous theory coined as the "simple man's model" or the "three-step model" \cite{LBI1994} -- which had its 25th anniversary in 2019 \cite{Symphony} -- dealt with the interaction of an electron with its parent nucleus sitting in a strong laser field environment, and elegantly described it according to classical and quantum processes. However, due to the fact that these laser pulses are highly coherent and contain huge numbers of photons, the description of the interaction in the strong field has so far been incomplete, because it treated the atomic system in a quantum way but the electromagnetic field in a classical way.

Now, in the description of the most relevant processes of ultra-intense laser-matter physics (such as high-harmonic generation, above threshold ionization, laser-induced electron diffraction, sequential and non-sequential multi-electron ionization, etc.) the quantum-fluctuation effects of the laser electric field, not even to mention the magnetic fields, are negligible. However, the quantum nature of the entire electromagnetic fields is always present in these processes, so a natural question arises: does this quantum nature exhibit itself? In which situations does it appear?

In the recent study \cite{Paris}, we have reported on the theoretical and experimental demonstration that intense laser-atom interactions may lead to the massive generation of highly non-classical states of light, one of the Holy Grails of the contemporary QS (see Fig. \ref{fig:paris}).

Such results have been obtained using the process of high-harmonic generation in atoms, in which large numbers of photons from a driving laser pulse of infrared frequency are up-converted into photons of higher frequencies in the extreme ultraviolet spectral range. The quantum electrodynamics theory formulated in this study, predicts that, if the initial state of the driving laser is coherent, it remains coherent but shifted in amplitude after interacting with the atomic medium.

Similarly, the quantum states of the harmonic modes become coherent with small coherent amplitudes. However, the quantum state of the laser pulse that drives the high-harmonic generation can be conditioned to account for this interaction, which transforms it into a, so-called, optical Schr\"odinger cat state. This state corresponds to a quantum superposition of two distinct coherent states of light: the initial state of the laser, and the coherent state reduced in amplitude that results from the interaction with the atoms.

We accessed the full quantum state of this laser pulse experimentally using quantum state tomography. To achieve this, the coherent amplitude of the light first needs to be reduced in a coherent way to only a few photons, on average, and then all of the quantum properties of the state can be measured.

The results of this study open the path for investigations towards the control of the non-classical states of ultra-intense light, and exploiting conditioning approaches on physical processes relevant to high-harmonic generation. This hopefully will link ultra-intense laser-matter physics and atto-science to quantum information science and quantum technologies in a novel and completely unexpected manner.

\subsection{Twistronics and moir\'e patterns}

\paragraph{\textit{Plethora of states in magic-angle graphene}}

In the recent years, graphene made another major splash in the headlines when scientists discovered that by simply rotating two layers of this material one on top of the other, this material could behave like a superconductor where electrical currents can flow without resistance. This new phase of matter turned out to appear only when the two graphene layers were twisted between each other at an angle of $\approx1.1$ degree (no more and no less) - the so-called magic angle - and it is always accompanied by enigmatic correlated insulator phases, similar to what is observed in mysterious cuprate high-temperature superconductors. 

\begin{figure}[t]
\centering
\includegraphics[width=7cm]{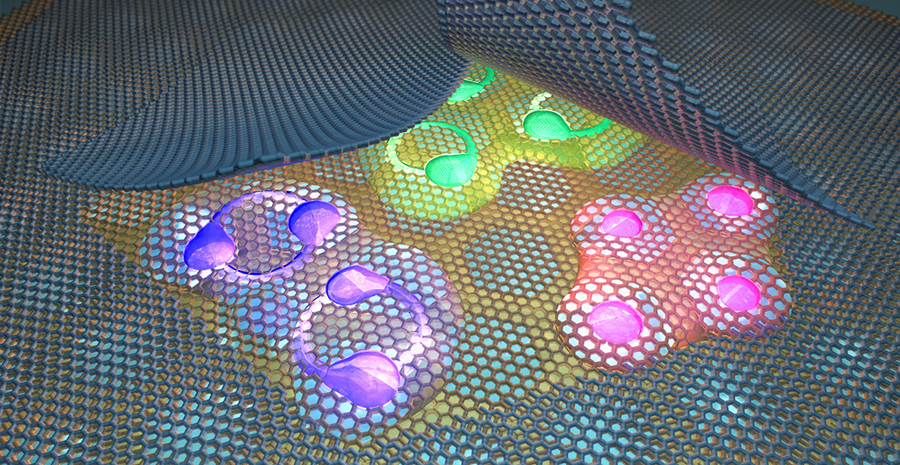}
\caption{Artistic illustration of the twisted bi-layer graphene and the different states of matter that have been discovered. Reprinted with permission of ICFO/ F. Vialla.} 
\label{fig:dima}       
\end{figure}

The group of D. Efetov at ICFO has succeeded in vastly improving the device quality of this setup, and in doing so, has stumbled upon something totally unexpected. The researchers were able to observe a zoo of previously unobserved superconducting and correlated states, in addition to an entirely new set of magnetic and topological states, opening a completely new realm of rich physics (see Fig. \ref{fig:dima}).

Room temperature superconductivity is the key to many technological goals such as efficient power transmission, frictionless trains, or even quantum computers, among others. When discovered more than 100 years ago, superconductivity was only plausible in materials cooled down to temperatures close to absolute zero. Then, in the late '80s, scientists discovered high temperature superconductors by using ceramic materials called cuprates. In spite of the difficulty of building superconductors and the need to apply extreme conditions (very strong magnetic fields) to study the material, the field took off as something of a holy grail among scientists based on this advance. Since last year, the excitement around this field has increased. The double mono-layers of carbon have captivated researchers because, in contrast to cuprates, their structural simplicity has become an excellent platform to explore the complex physics of superconductivity.

In the recent experiment led by D. Efetov \cite{Lu}, using a "tear and stack" van der Waals assembly technique, the scientists at ICFO were able to engineer two stacked monolayers of graphene, rotated by only 1.1 degree - the magic angle. They then used a mechanical cleaning process to squeeze out impurities and to release local strain between the layers. In doing this, they were able to obtain extremely clean twisted graphene bilayers with reduced disorder, resolving a multitude of fragile interaction effects.

By changing the electrical charge carrier density within the device with a nearby capacitor, they discovered that the material could be tuned from behaving as an insulator to behaving as a superconductor or even an exotic orbital magnet with non-trivial topological texture - a phase never observed before. What is even more astounding is the fact that the device entered a superconducting state at the lowest carrier densities ever reported for any superconductor, a completely new breakthrough in the field. To their surprise, they observed that the system seemed to be competing between many novel states. By tuning the carrier density within the lowest two flat moir\'e bands, the system showed correlated states and superconductivity alternately, together with exotic magnetism and band topology. They also noted that these states were very sensitive to the quality of the device, i.e. accuracy and homogeneity of the twist angle between two sheets of graphene layers. Last but not least, in this experiment, the researchers were also able to increase the superconducting transition temperature to above $3$ Kelvin, reaching record values which are twice as high as previously reported studies for magic-angle-graphene devices. 

What is exceptional about this approach is that graphene, a material that is typically poor on strongly interacting electron phenomena, now has been the enabling tool to provide access to this complex and exceptionally rich physics. So far, there is no unique theory that can explain the superconductivity in magic angle graphene at the microscopic level, however with this new discovery, it is clear that a new chance to unveil its origin has emerged.

\paragraph{\textit{Photonic moir\'e patterns}}

\begin{figure}[t]
\centering
\includegraphics[width=7cm]{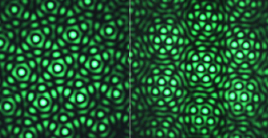}
\caption{moir\'e lattices created by superposition of two rotated hexagonal lattices. Reprinted with permission of ICFO/L. Torner.} 
\label{fig:lluis}       
\end{figure}

If you take two identical layers of semi-transparent material with the same structure, you put one on top of the other, rotate them and then look at them from above, hexagonal patterns start to emerge. These patterns are known as moir\'e patterns or moir\'e lattices.

moir\'e patterns appear often in every-day life applications such as art, textile industry, architecture, as well as image processing, metrology and interferometry. They are a matter of major current interest in science, since they can be produced using coupled graphene -hexagonal boron nitride monolayers, graphene -graphene layers and graphene quasicrystals on a silicon carbide surface and have proven to generate different states of matter upon rotating or twisting the layers to a certain angle, opening to a new realm of physics. Scientists at MIT found a new type of unconventional superconductivity in twisted bilayer graphene that forms a moir\'e lattice, and a team of ICFO researchers recently unveiled a new zoo of unobserved states in the same structure (see above). 

Now in a study published in Nature, Lluis Torner and collaborators have reported on the propagation of light in photonic moir\'e lattices, which, unlike their material counterparts, have readily controllable parameters and symmetry, allowing researchers to explore transitions between structures with fundamentally different geometries (periodic, general aperiodic and quasicrystalline - for the earlier theory paper see \cite{scirep}).  Note that both the theory and the experiments were done at the classical regime, with classical light, but the idea works for any kinds of waves, like matter waves, Bose-Einstein condensates, etc. 

The paper shows the creation of moir\'e lattices by superimposing two periodic patterns with either square or hexagonal primitive cells and tunable amplitudes and twist angle. Depending on the twist angle, a photonic moir\'e lattice may have different periodic (commensurable) structure or aperiodic (incommensurable) structure without translational periodicity. The angles at which a commensurable phase (periodicity) of a moir\'e lattice is achieved are determined by Pythagorean triples or by another Diophantine equation depending on the shape of the primitive cell. Changing the relative amplitudes of the sublattices allowed researchers to smoothly tune the shape of the lattice without affecting its rotational symmetry.

Then, by using commensurable and incommensurable moir\'e patterns, researchers observed for the first time the two-dimensional localization-delocalization transition of light. The used photonic moir\'e lattices can be readily constructed in practically any arbitrary configuration consistent with symmetry groups, thus allowing the creation of potentials that may not be easily produced in tunable form using material structures. Therefore, in addition to their direct application to the control of light patterns, the availability of photonic moir\'e patterns allows the study of phenomena relevant to other areas of physics, particularly to condensed matter, which are harder to explore directly.

\subsection{Quantum simulation for strongly correlated phases of matter}

\paragraph{\textit{Checkerboard insulator with semiconductor dipolar excitons}}

The Hubbard model constitutes one of the most celebrated theoretical frameworks of condensed-matter physics. It describes strongly correlated phases of interacting quantum particles confined in a lattice potential.In the last two decades the Hubbard Hamiltonian for bosons has been deeply scrutinised in the regime of short-range on-site interactions. On the other hand, extensions to longer-range interactions between neighbouring lattice sites have remained mostly elusive experimentally. Entering this regime constitutes a well identified research frontier where quantum matter phases can spontaneously break the lattice symmetry. In Ref. \cite{dubin} we unveil one of such phases, precisely the long-sought-after checkerboard solid. It is accessed by confining semiconductors dipolar excitons in a two-dimensional square lattice. The exciton checkerboard is signalled by a strongly minimised compressibility of the lattice sites at half-filling, in quantitative agreement with theoretical expectations. Our observations thus highlight that dipolar excitons enable controlled implementations of extended Bose-Hubbard Hamiltonians. 

\paragraph{\textit{Quantum simulation of interaction induced topological phases}}
 
In the last decades, topological insulators have attracted great interest and also have promising applications in topics such as metrology or quantum computation. These exotic materials go beyond the standard classification of phases of matter: they are insulating in their bulk, conducting on their edges, and characterized by a global topological invariant, in contrast to a local order parameter as in the conventional Ginzburg-Landau theory of phases of matter. The further discovery of a great variety of unexpected properties has led to an intensive investigation of topological phases of matter in many areas of quantum physics ranging from solid state and quantum chemistry to high-energy physics.

The research on topological insulators reached its first peak in 1997, when Altland and Zirnbauern classified all the possible topological phases of non-interacting systems by means of symmetry arguments. Even though this classification shed light on many novel features of quantum matter, the picture is still far from being complete. The list of exotic topological phases that escape the current classification is only growing, with a special focus on the discovery of interaction-induced topology in strongly correlated systems.

The study of interaction-induced topological phases is particularly challenging due to the high level of complexity encoded in interacting systems, which nowadays needs to be tackled through quantum algorithms working on classical processors. Nevertheless, the advent of a new generation of quantum simulators made of particles at ultracold temperatures represents reliable a rout towards a complete understanding and control of topological phases in interacting systems.

\begin{figure}[t]
\centering
\includegraphics[width=7cm]{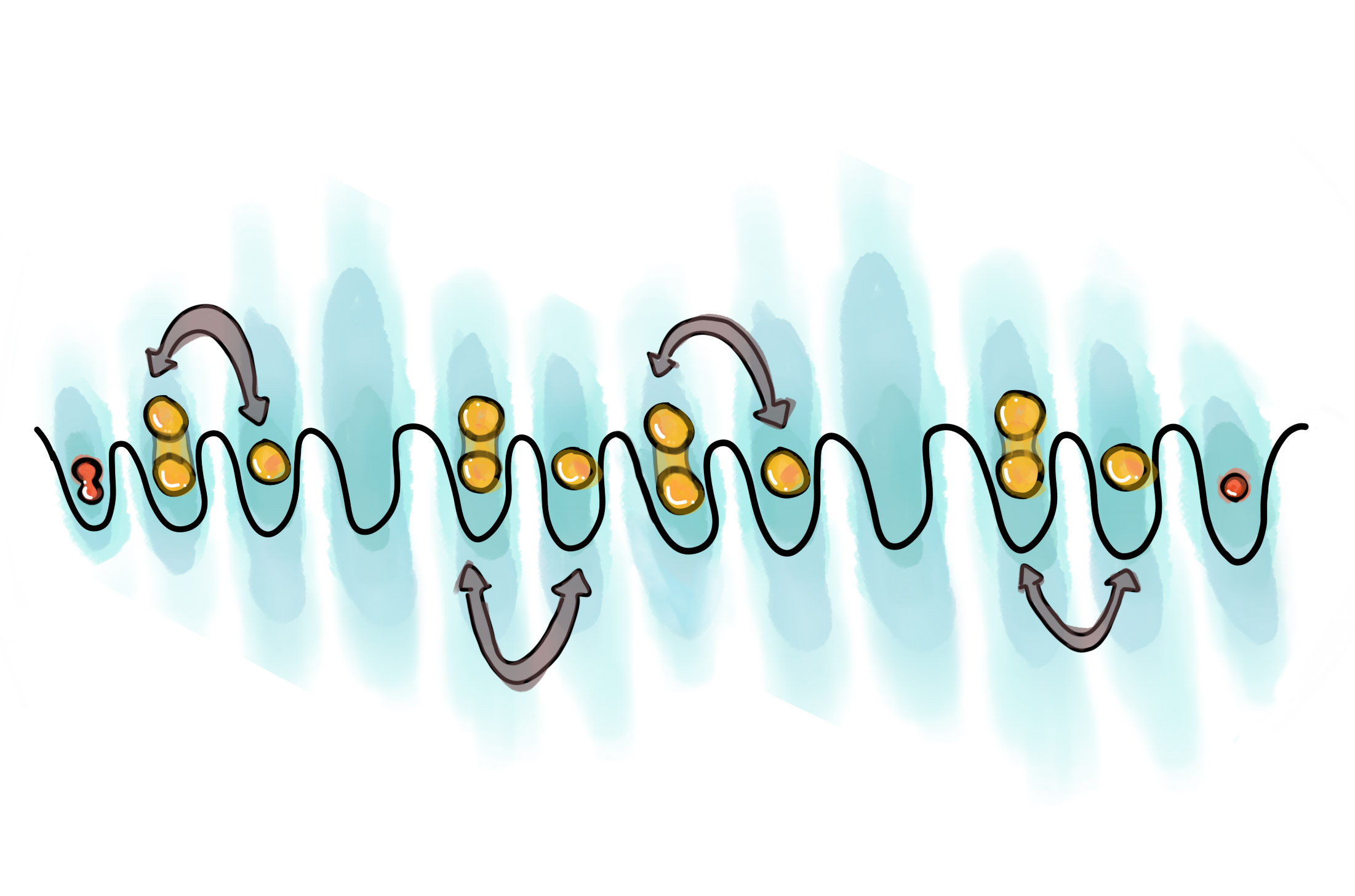}
\caption{The Haldane insulator is a symmetry protected topological phase which is found in the extended Bose Hubbard model at filling one. The topology extends to the critical point separating the Haldane insulator from a charge density wave. Reprinted with permission of ICFO/ J. Fraxanet.} 
\label{fig:joana}       
\end{figure}

In a recent article \cite{Joana}, the authors revealed that interacting processes can lead to topological properties persisting at the specific points separating two distinct phases, thus leading to the existence of topological quantum critical points (see Fig. \ref{fig:joana}). In addition, they also propose a realistic scheme based on magnetic dysprosium atoms whose long-range dipolar interaction enables the implementation of topological quantum critical points on a quantum simulator.

More specifically, they reported the presence of two distinct topological quantum critical points with localized edge states and gapless bulk excitations. The results show that the topological critical points separate two phases, one topologically protected and the other topologically trivial, both characterized by a string correlation function which denotes a similar type of long-range ordering. In both cases, the long-range order persists also at the topological critical points and explains the presence of localized edge states protected by a finite charge gap. Finally, they introduce a super-resolution quantum gas microscopy scheme for dipolar dysprosium atoms for the experimental study of topological quantum critical points using cold atoms in optical lattices.

\begin{figure}[t]
\centering
\includegraphics[width=7cm]{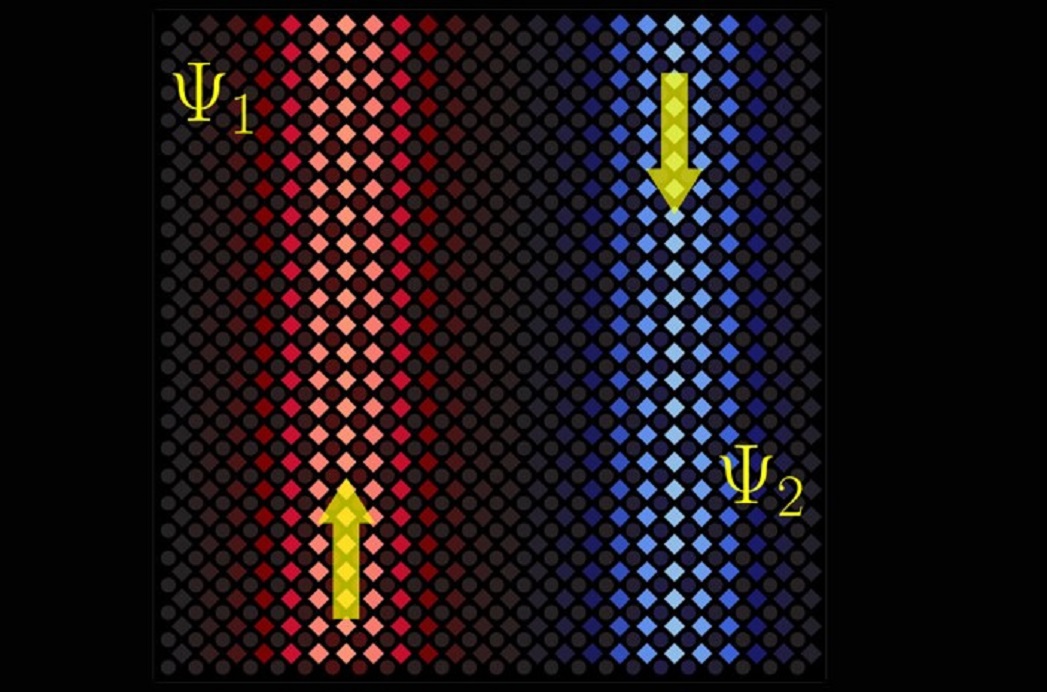}
\caption{Linear domain walls: a metastable self-consistent solution, in which the system develops two disconnected domain walls with ground state currents flowing in opposite directions. Reprinted with permission of ICFO/ S. Juli\`a.} 
\label{fig:sergi}       
\end{figure}



The quantum simulation of these exotic materials typically relies on the generation of artificial gauge fields. However, recent studies have shown that topological phases can also emerge from particle interactions. The latter mechanism leads to the concept of interaction-induced topological phases, in which topology is acquired through a spontaneous symmetry breaking process. The interplay of the spontaneous symmetry breaking with the global topological properties can lead to very interesting effects.

In another recent article \cite{Sergi}, they report how such interplay can lead to new strongly-correlated topological effects in a 2D material. They shown how interactions can localize particles in the insulating bulk, leading to self-trapped polarons. Moreover, they have also shown how the interacting nature of the topological insulator gives rise to domains in the bulk. Interestingly, the nontrivial topology associated to each domain leads to the appearance of protected conducting states in the bulk, localized at the domain boundaries (see Fig. \ref{fig:sergi}). Finally, they discuss the possibility of quantum simulating such phases with cold laser-excited Rydberg atoms in an optical lattice.

\subsection{Quantum simulation with Rydberg atoms}

The progress of this platform, led by the groups of Misha Lukin at Harvard and Antoine Browaeys at Institute d'Optique is truly amazing. Below, we include a lit of the most important applications and discoveries from Harvard-MIT: 

\paragraph{\textit{ Rydberg atoms for optimization algorithms}}

In the recent preprint \cite{Lukin2022}, the authors explain: "Realizing quantum speedup for practically relevant, computationally hard problems is a central challenge in quantum information science. Using Rydberg atom arrays with up to 289 qubits in two spatial dimensions, we experimentally investigate quantum algorithms for solving the Maximum Independent Set problem. We use a hardware-efficient encoding associated with Rydberg blockade, realize closed-loop optimization to test several variational algorithms, and subsequently apply them to systematically explore a class of graphs with programmable connectivity. We find the problem hardness is controlled by the solution degeneracy and number of local minima, and experimentally benchmark the quantum algorithm's performance against classical simulated annealing. On the hardest graphs, we observe a superlinear quantum speedup in finding exact solutions in the deep circuit regime and analyze its origins."

\paragraph{\textit{ Quantum spin liquids}}

In an earlier paper \cite{Lukin2021}, the group reported on quantum spin liquids, exotic phases of matter with topological order, that have been a major focus of explorations in physical science for the past several decades. As they explain: "Such phases feature long-range quantum entanglement that can potentially be exploited to realize robust quantum computation. We use a 219-atom programmable quantum simulator to probe quantum spin liquid states. In our approach, arrays of atoms are placed on the links of a kagome lattice and evolution under Rydberg blockade creates frustrated quantum states with no local order. The onset of a quantum spin liquid phase of the paradigmatic toric code type is detected by evaluating topological string operators that provide direct signatures of topological order and quantum correlations. Its properties are further revealed by using an atom array with nontrivial topology, representing a first step towards topological encoding. Our observations enable the controlled experimental exploration of topological quantum matter and protected quantum information processing."

\paragraph{\textit{Quantum scars}}

Perhaps the most impressive result is the control of quantum dynamics in Rydberg QS. In the already mentioned paper \cite{LukinScars}, the authors state that "controlling non-equilibrium quantum dynamics in many-body systems is an outstanding challenge as interactions typically lead to thermalization and a chaotic spreading throughout Hilbert space. We experimentally investigate non-equilibrium dynamics following rapid quenches in a many-body system composed of 3 to 200 strongly interacting qubits in one and two spatial dimensions. Using a programmable quantum simulator based on Rydberg atom arrays, we probe coherent revivals corresponding to quantum many-body scars. Remarkably, we discover that scar revivals can be stabilized by periodic driving, which generates a robust subharmonic response akin to discrete time-crystalline order. We map Hilbert space dynamics, geometry dependence, phase diagrams, and system-size dependence of this emergent phenomenon, demonstrating novel ways to steer entanglement dynamics in many-body systems and enabling potential applications in quantum information science. "

\paragraph{\textit{Quantum Phases of Matter on a 256-Atom Programmable Quantum Simulator}}

The title of this article \cite{Lukin256} is self explanatory. Still, it is worth to realize what the authors do: "Motivated by far-reaching applications ranging from quantum simulations of complex processes in physics and chemistry to quantum information processing, a broad effort is currently underway to build large-scale programmable quantum systems. Such systems provide unique insights into strongly correlated quantum matter, while at the same time enabling new methods for computation and metrology. Here, we demonstrate a programmable quantum simulator based on deterministically prepared two-dimensional arrays of neutral atoms, featuring strong interactions controlled via coherent atomic excitation into Rydberg states. Using this approach, we realize a quantum spin model with tunable interactions for system sizes ranging from 64 to 256 qubits. We benchmark the system by creating and characterizing high-fidelity antiferromagnetically ordered states, and demonstrate the universal properties of an Ising quantum phase transition in (2+1) dimensions. We then create and study several new quantum phases that arise from the interplay between interactions and coherent laser excitation, experimentally map the phase diagram, and investigate the role of quantum fluctuations. Offering a new lens into the study of complex quantum matter, these observations pave the way for investigations of exotic quantum phases, non-equilibrium entanglement dynamics, and hardware-efficient realization of quantum algorithms."

\section{Design, techniques and diagnostics of quantum simulators}
\label{sec:6}

In this Section we shortly mention the most important diagnostics and design tools that have been developed in the recent  years for QS, mostly with the field of atomic physics, providing interesting complementary methods for condensed matter physics.

\subsection{Single site and single particle resolution}
 
A quantum gas microscope is capable to detect single atoms in a single site of the (optical) lattice. In the first experiment of the Harvard group \cite{Bakr} (see also \cite{Sherson}),  the device was capable only to distinguish between zero/two or one bosonic atom at the site. Still, it offered an unprecedented possibility of measuring density-density correlations far beyond the possibilities of the standard condensed matter physics. These devices are nowadays capable to measure single particles in a single site of the lattice with spin resolution \cite{ReviewGross} for both bosons and fermions \cite{MicroscopeFermions}, and even more (for recent reviews see \cite{Kuhr}).

\subsection{Entanglement and topology characterization using random unitaries}
  
Another aspect that is extremely important to characterize the quality of QSs, is clearly the verification  of entanglement in the considered systems. This goes back to the theory and applications of entanglement witnesses (cf. \cite{LSA2017}), but more recently focuses on measurement of entanglement entropies of the reduced density matrices and entanglement Hamiltonians (logarithm of of the reduced density matrix) of partial blocks of a given system. In this respect, there is a spectacular progress led by Peter Zoller's group collaborating with various experimental teams. In a series of works they propose to apply random unitary operations to a part of the system (a block) and then perform local measurements there \cite{Elben,Dalmonte}. Experiments were realized with trapped ions \cite{Brydges}, while the method generalizes to detect topological invariants \cite{Hafezi}.

This is how the authors describe their ideas in a recent review \cite{Elben1}: 
"Increasingly sophisticated programmable quantum simulators and quantum computers are opening unprecedented opportunities for exploring and exploiting the properties of highly entangled complex quantum systems. The complexity of large quantum systems is the source of their power, but also makes them difficult to control precisely or characterize accurately using measured classical data. We review recently developed protocols for probing the properties of complex many-qubit systems using measurement schemes that are practical using today's quantum platforms. In all these protocols, a quantum state is repeatedly prepared and measured in a randomly chosen basis; then a classical computer processes the measurement outcomes to estimate the desired property. The randomization of the measurement procedure has distinct advantages; for example, a single data set can be employed multiple times to pursue a variety of applications, and imperfections in the measurements are mapped to a simplified noise model that can more easily be mitigated. We discuss a range of use cases that have already been realized in quantum devices, including Hamiltonian simulation tasks, probes of quantum chaos, measurements of nonlocal order parameters, and comparison of quantum states produced in distantly separated laboratories. By providing a workable method for translating a complex quantum state into a succinct classical representation that preserves a rich variety of relevant physical properties, the randomized measurement toolbox strengthens our ability to grasp and control the quantum world."

\subsection{ Experiment-friendly approaches for entanglement characterization}
  
Our approach for the characterization of entanglement and quantum correlations was different from that of Peter Zoller's group. Instead of looking for methods of measuring Renyi entropies and more, we focused on the characterization of quantum correlations via measurement of experimentally friendly low moments of local observables. In the first paper \cite{Jordi}, we characterize all permutationally invariant Bell inequalities in many body (party) systems with two observables per party with two outcomes. Our results were immediately tested in experiments with spinor Bose Einstein condensates and more.

Entanglement is one of the most characteristic phenomena of quantum physics. The simplest and most studied form of entanglement is the bipartite case, in which two subsystems form a quantum composite (e.g., two entangled particles). However, systems with more than two particles can exhibit entanglement in a whole plethora of ways, presenting a much richer and challenging case. Contrary to the bipartite case, multipartite entanglement admits a hierarchy of definitions depending on the strength of the correlations between the subsystems forming the quantum system. Therefore, as the quantum system becomes larger, it becomes more challenging to characterize it.

\begin{figure}[t]
\centering
\includegraphics[width=7cm]{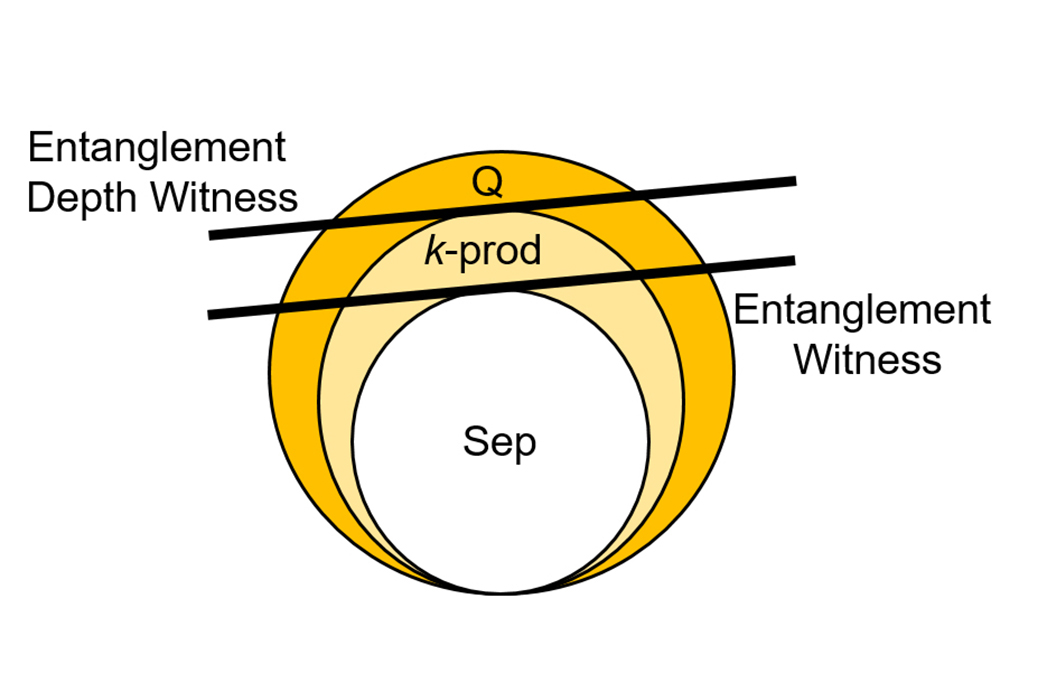}
\caption{Schematic representation of an Entanglement Depth Witness. Figure's author: J. Tura.} 
\label{fig:jordi}       
\end{figure}

Interestingly, the degree of violation of our inequalities gives direct information about the nature of entanglement inherent in the considered many body state \cite{Tura,Jordi}, and provides a method to classify the different degrees of multipartite entanglement in a computationally and experimentally tractable way (see Fig. \ref{fig:jordi}).

In this work, the researchers present a method to derive Device-Independent Witnesses of Entanglement Depth (DIWEDs) from Bell inequalities. Such DIWEDs quantify the strength of entanglement on a quantum many-body system by observing Bell non-local correlations. The difficultly of this approach lies in how one can derive these DIWEDs from Bell inequalities. The researchers have been able to come up with an elegant solution that reduces the issue to an efficient optimization problem. In particular, their methodology finds the maximal amount of Bell non-local correlations that can be achieved given any quantum system that has at most $k$ particles entangled. This provides a hierarchy of certification bounds, since such values can only be surpassed by quantum systems that have more than $k$ particles entangled. In addition, they have also been able to show how the DIWEDs can be rewritten in terms of collective measurements, and then apply these DIWEDs on existing experimental data in order to certify an entanglement depth of $15$ particles in a Bose-Einstein condensate of $480$ particles. 

Recently, we have continued to work on this approach to quantum correlations, and we were able to formulate Bell correlation and entanglement tests entirely based on accessible experimental or simulation data \cite{Guillem1,Guillem2}.
  
\subsection{ Experiment-friendly approaches for characterization of topology}  
  
We quote here only two examples of recent experimental work. More experiment-friendly approaches will be mentioned in the context of synthetic dimensions below. The first example comes from yet another platform for QS: polaritons, the second from photonics. 
 
Topological insulators attracted much interest in the last decades. These exotic materials are characterized by a global topological invariant, an integer, that cannot be deformed by local perturbations such as disorder or interactions. This robustness makes them ideal candidates for applications in metrology or quantum computation. For example, materials displaying the integer quantum Hall effect yield extremely robust plateaux in their transverse conductivity. Such robustness comes from the celebrated bulk-edge correspondence, which dictates that the topological invariant of the bulk is equal to the number of protected conducting edge states.

A distinct situation arises in 2D crystals like graphene, possessing both time reversal and chiral symmetry. Such systems do not have a gap. Nevertheless, they can present edge states which are robust against perturbations respecting the symmetry of the system. These states can be linked to a topological invariant defined over reduced one-dimensional subspaces of their Brillouin zone.

\begin{figure}[t]
\centering
\includegraphics[width=7cm]{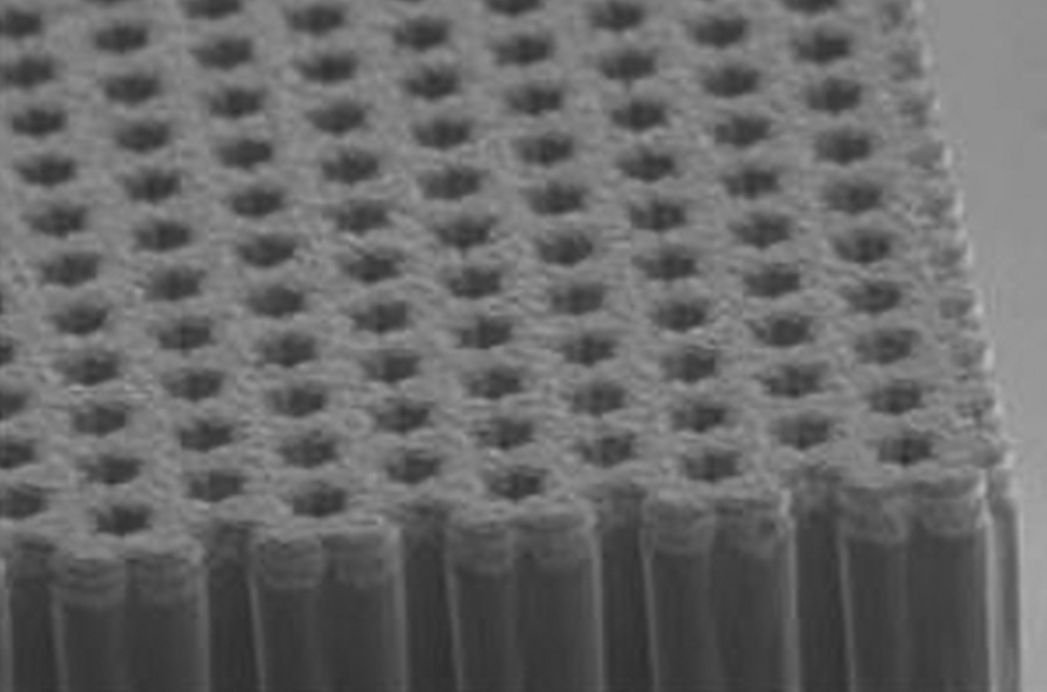}
\caption{Scanning electron microscopy image of a honeycomb lattice of coupled micropillars. Source: ICFO webpage news} 
\label{fig:alex}       
\end{figure}

In a recent article published in Physical Review Letters \cite{Alex} and highlighted as an Editor's Suggestion, the authors reported the measurement of such topological invariants in artificial graphene. They demonstrated a novel scheme based on a hybrid position- and momentum-space measurement to directly access these 1D topological invariants in lattices of semiconductor microcavities confining exciton-polaritons. They showed that such technique can be applied both to normal and strained graphene. This work opens, in our opinion, the door to a systematic study of such systems in the presence of disorder or interactions (see Fig. \ref{fig:alex}).

Yet another novel way of detecting topological properties relates to photonic systems. Research on topological insulators is moving at fast speed, promising a broad spectrum of applications ranging from metrology to quantum computing. Topological insulators are a new phase of matter in which the bulk of the material is an insulator, but its edges conduct electricity through what is called a "topological protected" edge states, that are a direct manifestation of the nontrivial topology hidden in their band structure.

To understand the effects of topology, physicists are working simultaneously on a plethora of experimental architectures. Among these, quantum walks are powerful models where topological phases of matter can be simulated in static and out-of-equilibrium scenarios. Most of the quantum walk architectures built so far generated one dimensional processes. However, there are currently efforts on increasing the dimensionality of these platforms to investigate the broad range of topological phenomena that exist in 2D and 3D.

\begin{figure}[t]
\centering
\includegraphics[width=7cm]{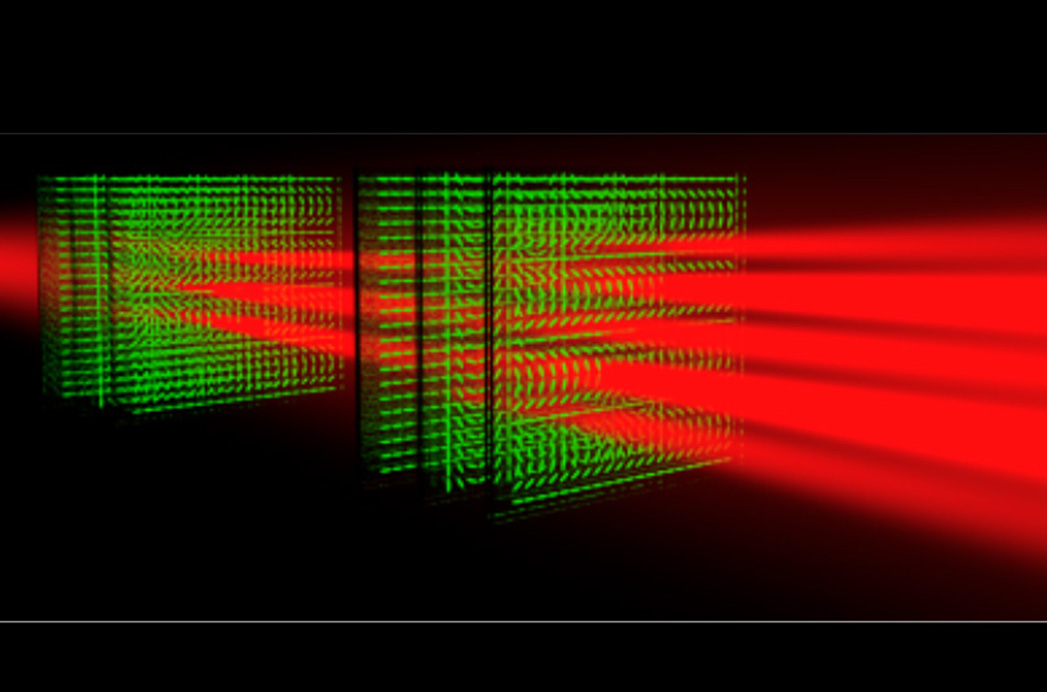}
\caption{A collimated beam crosses a sequence of liquid-crystal (LC) devices. Different LC patterns implement coin rotations and spin-dependent walker discrete translations. Reprinted with permission of SLAM research group (Naples).} 
\label{fig:cardano}       
\end{figure}

In a study recently published in Optica \cite{Cardano}, researchers reported on a novel photonic platform, capable of producing quantum walks of structured photons in two spatial dimensions, and exhibiting quantum Hall behavior. In their study, the researchers report on the realization of a photonic platform generating a quantum walk on a two-dimensional square lattice, that emulates a periodically-driven quantum Hall insulator. The apparatus consists of cascaded liquid crystal slabs, patterned to give polarization-dependent kicks to the impinging photons. Suitable combinations of these plates allow to manipulate dynamically the evolution of a light beam, realizing a quantum walk between light spatial modes carrying a variable amount of transverse momentum. The authors demonstrate the non-trivial topological character of their photonic system by directly reading out the anomalous displacement of an optical wavepacket when a constant force is introduced in the system (see Fig. \ref{fig:cardano}).

The simulation of other condensed matter systems, the investigation of the evolution of quantum light and the study of dynamical phase transitions are among the possible paths that the researchers of the study intend to explore in the next future.

\subsection{Synthetic dimensions}
 
\begin{figure}[t]
\centering
\includegraphics[width=\textwidth]{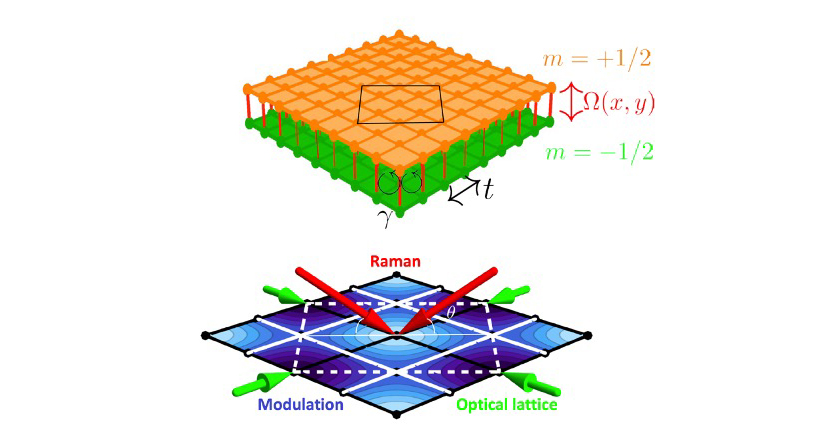}
\caption{Twistronics without a twist. Upper panel: synthetic bilayer of (pi-flux) square lattices with periodicity equal to 4 sites both in x and y axis. Such modulation of $\Omega{x,y}$ supports Dirac cones in single particle dispersion. Lower panel: Spatially modulated Raman coupling between the two layers mimics moir\'e patterns. Figure's author: T.Salamon/L.Tarruell.} 
\label{fig:tymek}       
\end{figure}

The idea of synthetic dimensions was formulated by us \cite{Boada}, and was independently discovered in condensed matter physics. In AMO physics it gained particular interest when it was formulated directly in the contexts of using internal atomic states for synthetic dimension, and employing designed Raman couplings to create synthetic gauge fields \cite{CeliPRL} - indeed the first top level experiments came out few months after our paper. Amazingly, the synthetic dimensions approach allows, for instance,  to measure Chern invariants in extremely narrow quasi-2D strips (see \cite{Mugel} and references therein).
 
Perhaps the most creative use of the idea of the synthetic dimension concerns, however, twistronics. The discovery of flat-bands in magic angle graphene has established a whole new concept - twistronics - which allows to induce strong correlation effects, anomalous superconductivity, magnetism and topology. The creation of this entirely novel field opens entirely new possibilities as well as previously unreachable degrees of manipulation and generation of the quantum many body states. Not only has it far reaching implications in the fundamental understanding of these exotic phases, but it is also likely to make far-reaching changes in technology, for instance in the fields of quantum metrology and sensing. The online symposium "Emergent phenomena in moir\'e materials", organized by ICFO and MIT illustrated perfectly the enormous interest and progress in this area.

The interesting phenomenology of twisted materials is apparently related to the formation of moir\'e patterns around small twist angles. Some of these twisting angles, the so-called magic angles, lead to vast band flattening already at the single-particle level. The geometrical moir\'e patterns induce spatially varying interlayer couplings that are behind the strong modification of the band structure. Emulating this physics beyond materials research can help in the identification of key minimal ingredients that give rise to the phenomenology of Twisted Bi-layer Graphene (TBLG), while also providing additional microscopic control. Photonic systems, for example, are well suited to explore this physics at the single-particle level as shown in a recent paper by Lluis Torner and his collaborators \cite{Torner} (see above).

Ultracold atoms in optical lattices are, in principle, a very promising platform to experimentally explore also the corresponding emerging many body phenomena. In a recent letter \cite{tymek}, we have proposed an atomic quantum simulator of twisted bilayers without actual physical twisting between layers. The idea is to elegantly employ synthetic dimensions. Atoms are located in a single 2D lattice, but can occupy two internal states, mimicking two layers. Coupling between the layers is realized by Raman transitions that are spatially modulated so as to mimic the action of twists, or in general arbitrary desired moir\'e  patterns. The advantages of this scheme are clear:
\begin{itemize}
\item Control of interlayer tunneling strength over wide ranges.
\item Control of atom-atom interaction, so that the ratio of kinetic-to-interaction energy can be tuned over wide ranges.
\item Magic angle physics appears at larger angles with smaller moir\'e supercells, implying less fluctuations of the twisting angle.
\end{itemize}

\subsection{Methods for verification and certification of quantum simulators}

To verify, validate and certify QSs,  we need {\bf theoretical methods} to design {\bf experimental strategies}  that are at least capable to describe qualitatively and even more importantly quantitatively the phenomena/systems in question. In this section we list the ones that have been developing particularly rapidly in the recent years:

\paragraph{\textit{Tensor networks}}

Tensor networks (TN) are novel numerical methods suitable for studies of strongly correlated systems. For lattice models, they involve tensors defined at each lattice site that depend on the physical variables at this site (spin state, bosonic state, ...). These tensors live also in auxiliary (bond) spaces, through which they are connected (via index contraction) to other sites. In one dimension this approach goes back to the Density Matrix Renormalization Group (DMRG) method of Steve R. White \cite{white}. More contemporary versions are termed Matrix Produc States (MPS), or in time dependent version Time-Evolving Block Decimation (TEBD). While standard TN are perfectly describing states that fulfill the, so called, entanglement area law, stronger forms of entanglement are captured by the Multiscale Entanglement Renormalization Ansatz (MERA). The analog of MPS in 2D and higher dimensions is termed Projected Entangled Pair States (PEPS). There are many review and handbook about TN mehtods, published recently; we refer here to our own book, published with open access in Lecture Notes in Physics \cite{shiju}.

\paragraph{\textit{Exact diagonalization}}

Obviously, exact diagonalization (ED) is limited to not-too-large systems, but is indispensable in some cases, such as for instance the studies of Many Body Localization (MBL). A state-of-art of ED is well represented by our recent study of disordered spin systems that exhibit MBL \cite{sierant}. 

\paragraph{\textit{Polynomial relaxations based on semi-definite programming}}

Many problems of interest are encoded in the ground state of an interacting many-bod system, classical or quantum. In this context, a quantum simulator provides a physical approximation to this unknown ground state. The measured energy therefore constitutes an upper bound to the ground-state energy. However, there is no certificate that the obtained solution is close to the unknown ground-state energy, which may significantly differ from the obtained value, for instance because of the presence of local minima. A complementary approach to verify the quality of the obtained upper bound is to construct methods that provide lower bounds to the ground-state energy. If the lower and upper bounds are close, one certifies that the approximation is close to the unknown solution. In the ideal case that the bounds coincide, one even concludes that the ground state has been reached.

A method to derive lower bounds to ground-state energy problem is given by hierarchies of relaxations to polynomial optimization problems based on semi-definite programming (SDP). The method was introduced by Lasserre~\cite{Lasserre} and Parrillo~\cite{Parrilo} for the classical case of commuting variables, and generalised to the quantum case of non-commuting operators by Navascu\'es, Pironio and Ac\'\i n~\cite{NPA,PNA}, see also~\cite{qmoment}. Without entering into the details of the formalism, in the classical case, these methods can be used when dealing with polynomial optimization problem as follows:
\begin{eqnarray}
&&\min_{\vec x} p(\vec x)\\
&&\text{such that}\nonumber\\
&&g_j(\vec x)\geq 0, \quad j=1,\ldots,m.\nonumber
\end{eqnarray}
Here $p$ and $g_j$ are polynomials over some variables $\vec x=(x_1,\ldots,x_n)$. The form of the quantum problem is basically the same, reading
\begin{eqnarray}
&&\min_{\ket\psi,\vec X} \langle\psi | p(\vec X)|\psi\rangle\\
&&\text{such that}\nonumber\\
&&g_j(\vec X)\geq 0, \quad j=1,\ldots,m.\nonumber
\end{eqnarray}
Now, polynomials are defined over some operators $\vec X=(X_1,\ldots,X_n)$ and inequalities should be understood as operator inequalities. We denote the solution to these problems by $p^*$. Many classical and quantum ground-state problems can be written in this polynomial form, for instance, as polynomials of (i) two-value variables for classical spins, (ii) pauli matrices for quantum spins or (iii) creation and anihilation operators for fermions.

Without entering into the details, the previous methods provide a hierarchy of lower bounds to the searched solution $p_1\leq p_2\leq p_3\ldots\leq p^*$. The interesting property is that each lower bound can be computed by semi-definite programming (SDP), a standard convex optimisation method~\cite{sdpref} that is not affected by local minima. The complexity of the SDP problem increases with the so-called level of the hierarchy. Under mild conditions, the hierarchy is proven to asymptotically converge, $p_\infty\rightarrow p^*$. In our context, the method provides a sequence of improving lower bounds to the ground-state energy that can be used to benchmark the outputs of variational methods or quantum simulators. 

SDP relaxations have been used in many different systems, for instance in quantum chemistry~\cite{Mazziotti,Nakata}, quantum spin systems~\cite{barthel,plenio} or, more recently, to benchmark the outputs of a D-Wave quantum annealer~\cite{baccari}.

\paragraph{\textit{Machine learning}}

\begin{figure}[t]
\centering
\includegraphics[width=6cm]{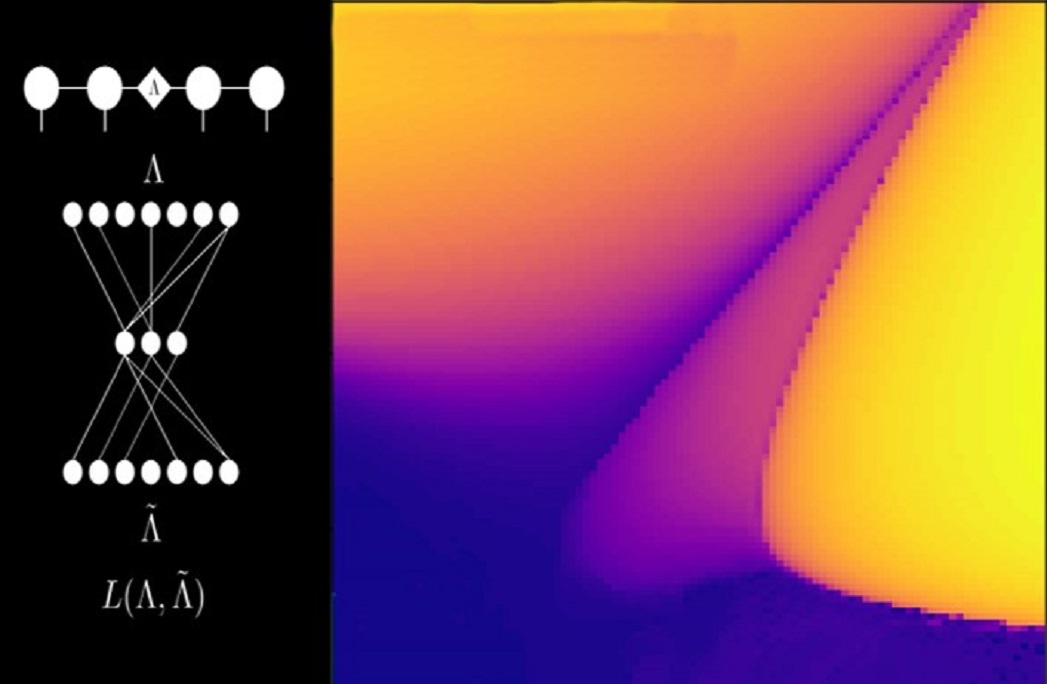}
\caption{Unsupervised Machine Learning in Physics. Figure's author: K.Kottmann.} 
\label{fig:bintz}       
\end{figure}

Machine learning (ML) and Artificial Intelligence (AI) can be found everywhere nowadays: from science through technology to politics and marketing. For the study of applications in physics and chemistry we recommend the recent overview lecture by the late Naftali Tishby on YouTube \cite{Tishby1}, the recent review \cite{Tishby2},  and our new book \cite{MLbook}. For the case of QSs, classical or quantum ML can be used to recognize quantum phases, quantum dynamics, and more, from data. The data can be generated in experiment or numerical simulation. A nice example of usefulness of ML to quantum many body physics is described in our recent Letter \cite{bintz}. 

Machine Learning (ML) has the main goal of analyzing and interpreting data structures and patterns in order to learn from them, reason and carry out a decision-making task that is completely independent from human reasoning and engagement. Even though this field of study started in the mid 1900s, recent developments in the area have revolutionized the way how we process and find correlations in complex data.

Contrary to supervised learning, unsupervised learning seeks to discover patterns or classify information in large data sets into categories without prior knowledge. That is, it does not have labeled outputs, which means that it basically infers the natural structure that a dataset may have and extracts categorized information from it. This learning has proved to be very efficient for identifying phases and phase transitions of many-body systems. In our Letter, we report on a method that uses an unsupervised machine learning technique based on anomaly detection to automatically map out the phase diagram of a quantum many body system given unlabeled data.

The following example is very illustrative of what has been achieved. In machine learning the most common and known classification task example is to discriminate, for instance, images of cats and dogs. In  particular, anomaly detection handles the classification task of discriminating dogs and everything that is not a dog, approaching the system is an entirely different perspective. The idea is to train a special neural network called an autoencoder to efficiently compress and reproduce images of dogs. If the network is later fed with images of cats, the network does not know how to efficiently compress the features of the cat image and it is possible to tell from the higher reconstruction loss that it is not a dog. In \cite{bintz}, the authors use this method in the context of quantum many-body systems. The images become observables, wave-functions or entanglement properties and the classes dogs and cats become different quantum phases. The model that they look at is the extended Bose-Hubbard model, which offers four different phases in the parameter space of interest. Since the researchers do not know the phases in their task a priori, they start by defining a region around the origin of the phase diagram as their starting point to train. Already from there, they are clearly capable of mapping the system in one training iteration, where all four phases of the system are easily distinguishable.

Finally, using this approach they have been able to reveal a phase-separated state between a supersolid and a superfluid, which appears in the system in addition to the standard superfluid, Mott insulator, Haldane-insulating, and density wave phases. The discovery of this new feature in this phase diagram is rather surprising provided that this model has been the investigated thoroughly for the past 20 years. This work is one of the first examples in which a machine detects a previously unknown phenomenon in a quantum many-body context. For a detailed analysis of this phenomenon and more, see the following Phys. Rev. B paper \cite{PRBBintz}.

\section{Conclusions}
\label{sec:7}

The area of quantum simulators is in any sense one of the most beautiful areas of contemporary physics, melting together in a common pot all possible branches and genres of physics and not only physics. The conclusion that we always propose is: Enjoy physics \& beyond!

In particular,  we at ICFO enjoy going beyond to try to interpret quantum mechanical processes, and more specifically quantum randomness in contemporary avantgarde music. We try indeed to incorporate quantum random processes, using the genuine quantum random number generators, provided to us by QUSIDE \cite{quside}. The highlight of our approach was the nearly one hour long concert "Interpreting Quantum Randomness" at the famous SONAR Festival 2021 \cite{sonar}.

Finally, we would like to end this essay, by commenting on relation between quantum computing and quantum simulation. Clearly, we bet on the latter. But, we are sure that  future of quantum computing is also bright; we expect it would be integrated with classical computing, combined in a symbiotic way with High Performance Computing (HPC) approaches. The distinction between quantum computing and simulation will probably become very diffuse, if not vanishing. The theoretical effort of computer scientists and theoretical physicists will play a pivotal role in this processes (cf. \cite{Zoltan}).

\begin{acknowledgement} We acknowledge Antonio Ac\'in and Leticia Taruell for their contributions to the manuscript. Over several years, we have been publishing mini-reviews of our "achievements" on the ICFO Web Page in the News section \cite{ICFO-web}. These text were edited with the help of Alina Hirschmann, whose irreplaceable help is highly appreciated.  We have adopted here some of these mini-reviews, mostly of the papers from our groups, but we also acknowledge courtesy of Adrian Bachtold, Alexandre Dauphin, Dima Efetov and Lluis Torner for being able to use their texts. JF, TS and ML acknowledge support from: ERC AdG NOQIA; Agencia Estatal de Investigaci\'on (R\&D project CEX2019-000910-S, funded by MCIN / AEI / 10.13039 / 501100011033, Plan National FIDEUA PID2019-106901GB-I00, FPI, QUANTERA MAQS PCI2019-111828-2, Proyectos de I+D+I "Retos Colaboraci\'on" QUSPIN RTC2019-007196-7);  Fundaci\'o Cellex; Fundaci\'o Mir-Puig; Generalitat de Catalunya through the European Social Fund FEDER and CERCA program (AGAUR Grant No. 2017 SGR 134, QuantumCAT \ U16-011424, co-funded by ERDF Operational Program of Catalonia 2014-2020); EU Horizon 2020 FET-OPEN OPTOlogic (Grant No 899794); National Science Centre, Poland (Symfonia Grant No. 2016/20/W/ST4/00314); European Union's Horizon 2020 research and innovation programme under the Marie-Sko\l dowska-Curie grant agreement No 101029393 (STREDCH) and No 847648  ("La Caixa" Junior Leaders fellowships ID100010434: LCF/BQ/PI19/11690013, LCF/BQ/PI20/11760031,  LCF/BQ/PR20/11770012, LCF/BQ/PR21/11840013). 
\end{acknowledgement}


\bibliographystyle{abbrv}
\bibliography{main.bib}

\end{document}